\documentclass[iop]{emulateapj}
\usepackage{apjfonts}
\usepackage{graphicx}
\usepackage{amsmath}
\usepackage{natbib}
\usepackage{IEEEtrantools}

\begin{document}

\title{Prospects for Characterizing Host Stars of the Planetary System Detections Predicted for the Korean Microlensing Telescope Network}
\author{
{Calen B.~Henderson}\altaffilmark{1}
}
\altaffiltext{1}{Department of Astronomy, The Ohio State University, 140 W.\ 18th Ave., Columbus, OH 43210, USA}
\email{henderson@astronomy.ohio-state.edu}


\shorttitle{Host Star Characterization Prospects for Predicted KMTNet Planet Detections}
\shortauthors{Henderson}

\begin{abstract}

I investigate the possibility of constraining the flux of the lens (i.e., host star) for the types of planetary systems the Korean Microlensing Telescope Network is predicted to find.
I examine the potential to obtain lens flux measurements by 1) imaging a lens once it is spatially resolved from the source, 2) measuring the elongation of the point spread function of the microlensing target (lens+source) when the lens and source are still unresolved, and 3) taking prompt follow-up photometry.
In each case I simulate observing programs for a representative example of current ground-based adaptive optics (AO) facilities (specifically NACO on VLT), future ground-based AO facilities (GMTIFS on GMT), and future space telescopes (NIRCAM on $JWST$).
Given the predicted distribution of relative lens-source proper motions, I find that the lens flux could be measured to a precision of $\sigma_{H_{\ell}} \leq 0.1$ for $\gtrsim$60$\%$ of planet detections $\geq$5 years after each microlensing event, for a simulated observing program using GMT that images resolved lenses.
NIRCAM on $JWST$ would be able to carry out equivalently high-precision measurements for $\sim$28$\%$ of events $\Delta t$ = 10 years after each event by imaging resolved lenses.
I also explore the effects various blend components would have on the mass derived from prompt follow-up photometry, including companions to the lens, companions to the source, and unassociated interloping stars.
I find that undetected blend stars would cause catastrophic failures (i.e., $>$50$\%$ fractional uncertainty in the inferred lens mass) for $\lesssim$(16$\cdot f_{\rm bin})\%$ of planet detections, where $f_{\rm bin}$ is the binary fraction, with the majority of these failures occurring for host stars with mass $\lesssim$0.3$M_{\odot}$.

\end{abstract}

\section{Introduction} \label{sec:intro}

Microlensing is an indispensable tool for understanding exoplanet demographics due to its unique sensitivity to low-mass planets separated from their host stars by a few AU or more.
This is underscored by the fact that this region roughly corresponds to the location of the snow line in protoplanetary disks, beyond which a higher surface density of solid material is thought to facilitate the growth of more massive protoplanets on shorter formation time scales (\citealt{lissauer1987,ida2005,kennedy2008}).

The current OGLE-IV \citep{udalski2003} and MOA-II \citep{bond2001,sumi2003} microlensing surveys collectively detect $\sim$15 planets per year.
However, converting the routinely measured mass ratio $q$ of the lens system (planet and host star) and the instantaneous projected angular separation $s$ into planet mass $M_{p}$ and instantaneous projected semimajor axis $a_{\bot}$ is difficult and requires additional information beyond the standard microlensing light curve.
Efficiently doing so will be all the more important due to the influx of data from the Korean Microlensing Telescope Network (KMTNet) \citep{kim2010,kim2011,kappler2012,poteet2012,atwood2012}, a next-generation network of microlensing survey telescopes that is predicted to increase the annual microlensing planet detection rate by a factor of $\sim$5 (\citealt{sexypants2014a}, hereafter H2014a).
There are two primary methods by which to determine $M_{p}$ and $a_{\bot}$ with minimal model dependence.

The first is by determining the microlens parallax $\pi_{\rm E}$, which can be measured from the distortion in the observed light curve due to the acceleration of the Earth relative to the light curve expected for a constant velocity \citep{gould1994,hardy1995,gould2009}.
If, for a given event, this resulting asymmetry as well as the angular size of the Einstein ring $\theta_{\rm E}$ can be measured, the latter typically by combining multiband photometry with a detection of finite-source effects, then the mass of the lens system can be derived from these two observables via
\begin{equation} \label{eq:mass_l}
   M_{\ell} = \frac{{\theta_{\rm E}}^{2}}{\kappa \pi_{\rm rel}},~\pi_{\rm rel} = \pi_{\rm E}\theta_{\rm E} = {\rm AU}(D_{\ell}^{-1} - D_{s}^{-1}),
\end{equation}
where $\pi_{\rm rel}$ is the relative lens-source parallax, $D_{\ell}$ and $D_{s}$ are the distances to the lens and source, respectively, and $\kappa \equiv 4G/(c^{2}{\rm AU}) = 8.144~{\rm mas}/M_{\odot}$.
This has hitherto been accomplished for 12 planetary systems, including a two-planet system \citep{han2013} and a circumbinary planet with mass twice that of Earth \citep{gould2014b}.
There are three different ways to measure $\pi_{\rm E}$, each with its own observational challenges.
Satellite parallax refers to when ground-based observatories and a space telescope are separated by a long spatial base line ($\sim$AU).
Orbital parallax can be measured for events with time scales that are a significant fraction of a year and requires good observational coverage.
Finally, terrestrial parallax occurs when multiple observatories at different longitudes monitor a high-magnification event simultaneously with extremely high cadences.
In all cases the stringent observational requirements indicate that the fraction of events for which it is possible to measure $\pi_{\rm E}$ is quite small.

The second is by constraining the flux of the primary lensing mass, the host star.
In the case that color information and finite-source effects provide $\theta_{\rm E}$, $M_{\ell}$ can be derived by measuring the lens flux, $F_{\ell}$, and applying a mass-luminosity relationship \citep{bennett2007}, given a value of the extinction toward the lens.
This method has been applied to only a few planetary microlensing events (e.g., \citealt{janczak2010,batista2014}), as it requires high-resolution follow-up photometry, typically in the near-infrared (NIR).
However, it does $not$ necessarily require waiting for the lens and source to be resolved.
In fact, there are several channels through which $F_{\ell}$ can be constrained: 1) imaging the lens after it is spatially resolved from the source, 2) inferring $F_{\ell}$ by measuring the elongation of the point spread function (PSF) of the unresolved microlensing target (lens+source) as the lens and source begin to separate, 3) promptly obtaining high-resolution follow-up photometry while the lens and source are unresolved, or 4) measuring a wavelength-dependent shift of the centroid of the unresolved microlensing target, stemming from the possibility that the lens and source have different colors.
There is an array of current and planned ground-based and space telescopes that will have the NIR detectors and diffraction-limited resolution $\theta_{\rm FWHM}$ necessary to employ these methods.

Here I present the results of simulated observing programs that explore the ability to constrain $F_{\ell}$ for predicted KMTNet planet detections.
I specifically investigate only items 1--3 listed above but note that measuring a color-dependent centroid shift is a useful tool and one that was successfully implemented for the first exoplanet discovered via microlensing \citep{bond2004,bennett2006}.
I give a review of the simulations of H2014a in \S \ref{sec:kmtnet_sims}.
In \S \ref{sec:high_res} I describe the specific facilities whose observational capabilities I consider.
I provide an overview of the practical implementation of each of these three techniques as well as my approximated methodology in \S \ref{sec:lens_flux_methods}.
In \S \ref{sec:results} I detail the results for each.
I then discuss the effects that contaminating flux from different types of blend stars would have in \S \ref{sec:contamination}.
Finally, in \S \ref{sec:discussion} I explain the implications my findings have for deriving masses of the planets that will be detected by KMTNet.

\section{Summary of KMTNet Simulations} \label{sec:kmtnet_sims}

The simulations of H2014a were designed to optimize the observing strategy for and predict the planet detection rates of KMTNet.
There are four primary components to their methodology:
\begin{itemize}
   \item using Galactic models to generate populations of lens and source stars with physical properties that match empirical constraints,
   \item populating each lens star with a single planetary companion and computing the magnification of the source star as a function of time,
   \item creating realistic observed light curves, and
   \item implementing a detection algorithm for each light curve.
\end{itemize}
Here I provide an overview of the details of each.

\subsection{Galactic Model} \label{sec:kmtnet_galmod}

H2014a use the luminosity function (LF) of \citet{holtzman1998} to obtain the absolute $I$-band magnitude of each source star, $M_{I,s}$.
Their Galactic bulge and disk density models come from \citet{han1995a} and \citet{han1995b}, respectively.
They draw $M_{\ell}$ from the mass function (MF) of \citet{gould2000}, which assumes that all main sequence stars in the range $1 < M_{\ell}/M_{\odot} < 8$ have become white dwarfs (WDs), in the range $8 < M_{\ell}/M_{\odot} < 40$ have become neutron stars (NSs), and in the range $40 < M_{\ell}/M_{\odot} < 100$ have become black holes (BHs).
All objects in the range $0.03 \leq M_{\ell}/M_{\odot} \leq 0.08$ are assumed to be brown dwarfs (BDs).
H2014a consider only main sequence stars as host stars of planetary systems, excluding BDs and remnants (WDs+NSs+BHs) from the underlying lens mass distribution.
The extinction map they use complements the $I$-band data of \citet{nataf2013} with the NIR map of \citet{majewski2011} and \citet{nidever2012} for the inner bulge.

\subsection{Microlensing Parameters} \label{sec:kmtnet_ulensparms}

There are four parameters that specify a microlensing event due to a single lensing mass.
The first is $t_{\rm 0}$, the time of closest approach of the source to the lens, which H2014a draw uniformly from a generic observing season.
Second is $u_{\rm 0}$, the angular distance of the closest approach of the source to the lens, normalized to $\theta_{\rm E}$.
H2014a set a maximum allowed impact parameter of 3 and draw its value uniformly.
The Einstein crossing time $t_{\rm E}$ is computed via
\begin{equation} \label{eq:t_E}
   t_{\rm E} \equiv \frac{\theta_{\rm E}}{\mu_{\rm rel}},
\end{equation}
where $\mu_{\rm rel}$ is the relative lens-source proper motion.
Last is $\rho$, the angular radius of the source star normalized to $\theta_{\rm E}$.

H2014a then populate each lens star with a planetary companion.
The mass ratio $q$ is given by
\begin{equation}
   q = \frac{M_{p}}{M_{\ell}}.
\end{equation}
H2014a assume a circular orbit for the planetary companion and compute $s$ via
\begin{equation}
   s = \frac{a}{R_{\rm E}}\sqrt{1-{\rm cos}^{2}\zeta},
\end{equation}
where $R_{\rm E}$ is the physical size of the Einstein ring radius and $\zeta$ is the angle between the plane of the sky and $a_{\bot}$ at the time of the microlensing event.
Finally, $\alpha$ gives the angle of the source trajectory relative to the star-planet binary axis and is drawn uniformly.
H2014a use these parameters to compute the magnification of the source due to the static binary lens system as a function of time.

\subsection{Light Curve Generation} \label{sec:kmtnet_lcgen}

H2014a then convert the magnification into an observed flux.
Their weather data for each KMTNet site come from \citet{peale1997} and they compute the brightness of the Moon using the prescription of \citet{krisciunas1991}.
H2014a determine the photon rate normalization and the flux measurement uncertainties for KMTNet by calibrating to OGLE-III photometry and scaling accordingly.

\subsection{Detection Algorithm} \label{sec:kmtnet_detalg}

Lastly, H2014a subject each simulated microlensing event to several detection criteria to determine if the planet is robustly detected.
First, the $\Delta\chi^{2}$ of the observed light curve from its error-weighted mean flux must be greater than 500.
Secondly, the light curve must have more than 100 data points and $t_{\rm 0}$ must fall within the time coverage of the light curve.
Finally, the $\Delta\chi^{2}$ of the light curve from a best-fit single-lens model must be greater than 160.
The detection rates are then normalized according to a modified version of the cool-planet mass function of \citet{cassan2012} that has been leveled-off at $M_{p} = 5M_{\oplus}$.

\section{High-resolution Facilities and Simulated Observational Programs} \label{sec:high_res}

\subsection{Current Ground-based Adaptive Optics} \label{sec:current_ground}

There are several large telescopes (>8m) with adaptive optics (AO) systems capable of achieving diffraction-limited resolution in the optical or NIR (see \S4 of \citealt{sexypants2014b} for an overview).
Of these, microlensing planet masses derived from $F_{\ell}$ have used $H$-band measurements made with NACO on VLT \citep{janczak2010} or NIRC2 on Keck \citep{batista2014}.
I utilize the former here as a representative example and simulate its observing capabilities.

At $\lambda = 1.66$ $\mu$m the full width at half maximum (FWHM) of a diffraction-limited image on the 8.2m VLT, given by 1.22$\lambda$/$D$, where D is the telescope aperture, is $\theta_{\rm FWHM,VLT}$ = 52.2 mas.\footnote{http://www.eso.org/sci/facilities/paranal/instruments/naco/doc.html}
I use their exposure time calculator (ETC)\footnote{http://www.eso.org/observing/etc/} to obtain the photon rate normalization, the sky background, and the scaling of the signal-to-noise ratio ($SNR$) with exposure time $t_{\rm exp}$.
For each method discussed in \S \ref{sec:lens_flux_methods} I simulate an observing program for each lens system H2014a predict KMTNet will detect, taking the aggregate sample to be characteristic of the types and variety of systems KMTNet will find.
My assumed input instrumental parameters for the simulated observing program are:
\begin{itemize}
   \item $H$-band observations, balancing PSF sharpness and resolution with sky background,
   \item an input spectrum of an M0V star (though the choice of template spectrum has little effect on the resulting $SNR$ or photon rate normalization),
   \item a laser guide star,
   \item the VIS dichroic, which has high efficiency for NIR observations,
   \item the S27 camera, which oversamples $H$-band slightly, and
   \item the FNS/HS instrument mode, which provides higher $SNR$ for fixed $t_{\rm exp}$ than does DCR/HD.
\end{itemize}

I set the minimum exposure time $t_{\rm exp,min}$ to be 20s, recommended for $H$-band, or whenever $SNR = 100$ is reached, and limit each observation to a maximum of 60 60s exposures.
Table \ref{tab:obs_parms} gives the parameters for the simulated observing program.

\begin{deluxetable*}{cccccccc}
\tablecaption{Parameters of Simulated Observing Programs}
\tablewidth{0pt}
\tablehead{
\colhead{Facility}                   &
\colhead{$\theta_{\rm FWHM}$}         &
\colhead{$t_{\rm exp,min}$}          &
\colhead{$t_{\rm exp,max}$}          &
\colhead{Collecting Area}            &
\colhead{Object Photon Rate$^{a}$}   &
\colhead{Background Photon Rate}     &
\colhead{Plate Scale}                \\
\colhead{}                           &
\colhead{[mas]}                      &
\colhead{[s]}                        &
\colhead{[s]}                        &
\colhead{[m$^{2}$]}                  &
\colhead{[e s$^{-1}$]}               &
\colhead{[e s$^{-1}$]}               &
\colhead{[mas pixel$^{-1}$]}
}
\startdata
NACO on VLT        &   52.2   &   20   &   3600   &   49.29   &   49.0   &   934    &   27.0   \\
GMTIFS on GMT      &   16     &   20   &   3600   &   368     &   366    &   239    &   5.0    \\
NIRCAM on $JWST$   &   68     &   11   &   3600   &   25      &   1290   &   4.77   &   31.7
\enddata
\tablenotetext{a}{For a point source with $H$ = 18.}
\label{tab:obs_parms}
\end{deluxetable*}

\subsection{Next-generation Ground-based Adaptive Optics} \label{sec:nextgen_ground}

There are currently three planned extremely large telescopes (>20m) that will each have an AO system and a NIR imager.
I select GMTIFS on GMT as my example with which to simulate an observing program because South Korea is a $10\%$ GMT partner, making the realization of such an endeavor as is proposed here all the more feasible and probable.

The 24.5m GMT will have a diffraction-limited resolution of $\theta_{\rm FWHM,GMT} = 16$ mas in $H$-band \citep{mcgregor2012} and a collecting area of 368m$^{2}$, $\sim$7.5 times that of VLT.\footnote{http://www.gmto.org/resources/}
To simulate an observing program on GMT I assume the same parameters as with VLT but I increase the photon rate normalization by the factor of 7.5 to account for the increase in aperture size and modify the sky background to include the increase in collecting area as well as the decrease in PSF area, arising from the smaller pixel size.
The parameters of the simulated observing program are listed in Table \ref{tab:obs_parms}.

\subsection{Next-generation Space-based Telescopes} \label{sec:nextgen_space}

\citet{bennett2006} used optical $HST$ observations to determine the mass of the first exoplanet discovered with microlensing \citep{bond2004}.
In looking forward, however, $JWST$ will provide the largest aperture yet in space at 6.5m and will use the NIR imager NIRCAM.
The bigger aperture provides a smaller diffraction limit than for $HST$ --- $\theta_{\rm FWHM,JWST} = 68$ mas for $JWST$'s $\lambda = 1.50$ $\mu$m short-wavelength filter.

I use the $JWST$ ETC\footnote{http://jwstetc.stsci.edu/etc/input/nircam/imaging/} with the following instrumental parameters:
\begin{itemize}
   \item the F150W filter, a good approximation of $H$-band,
   \item an M0V spectral distribution, and
   \item average zodiacal and thermal backgrounds.
\end{itemize}

I set $t_{\rm exp,min} =$ 11s (as suggested by the user's manual, accessed via the ETC page) or whenever $SNR = 100$ is reached and again set $t_{\rm exp,max} =$ 3600s.
Table \ref{tab:obs_parms} shows the parameters for the simulated observing program.

\section{Lens Flux Measurement Methods} \label{sec:lens_flux_methods}

\begin{figure*}
   \centerline{
      \includegraphics[width=9cm]{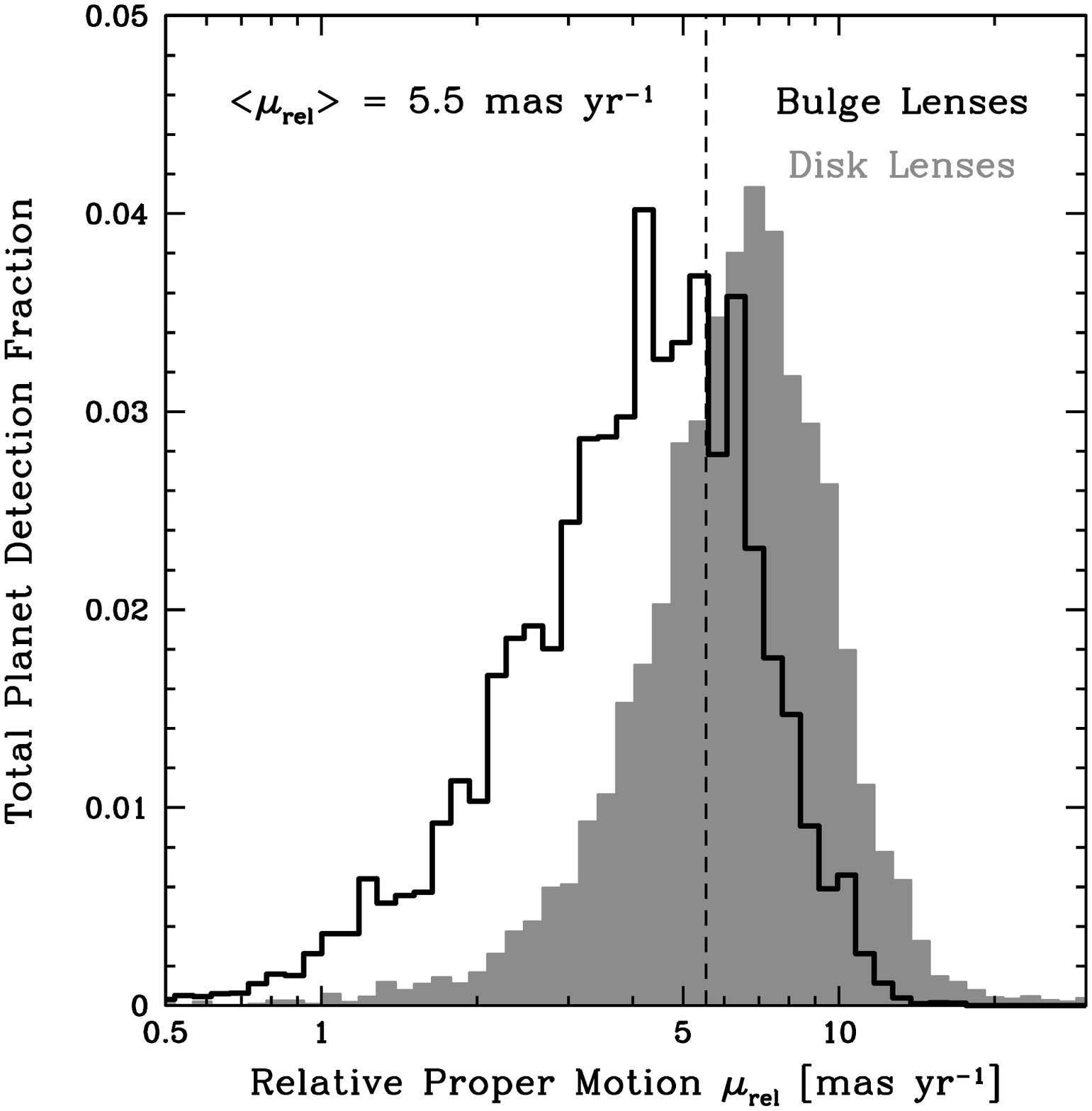}
      \includegraphics[width=9cm]{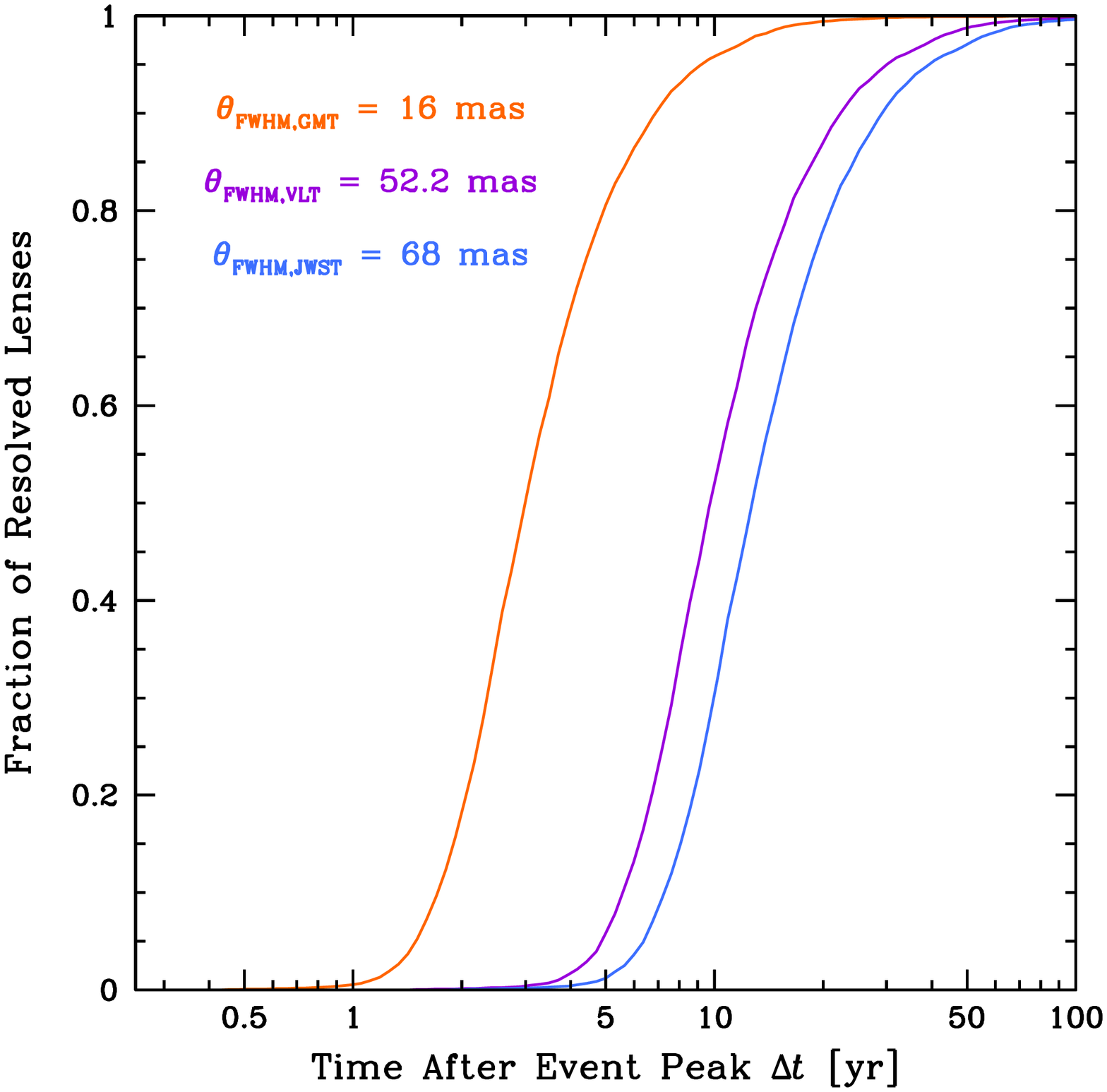}
   }
   \caption{
      \footnotesize{
         Distribution of relative lens-source proper motion $\mu_{\rm rel}$ for the planet detections predicted for KMTNet (left) and fraction of lenses that will then be resolved from the source as a function of $\Delta t$ for each facility (right).
         The fraction of microlensing events for which a given facility will be able to constrain $F_{\ell}$ depends sensitively on the fraction of events it can resolve a fixed time $\Delta t$ after each event.
         This population is similar for VLT and $JWST$, given their comparable diffraction-limited resolutions $\theta_{\rm FWHM}$, but is shifted toward significantly shorter values of $\Delta t$ for GMT.
      }
   }
   \label{fig:mu_res}
\end{figure*}

The feasibility of constraining $F_{\ell}$ for each technique explored here hinges on the relative lens-source proper motion, $\mu_{\rm rel}$.
The distribution of $\mu_{\rm rel}$ for the predicted KMTNet planet detections is shown in Figure \ref{fig:mu_res}.
It peaks at $\mu_{\rm rel} = 5.5$ mas yr$^{-1}$ and falls off more steeply toward larger values of $\mu_{\rm rel}$.
Microlensing events with lenses located in the Galactic disk generally have larger proper motions than do events arising from lenses in the Galactic bulge.
The efficacy of a given observational facility to constrain $F_{\ell}$ is set by the fraction of lens systems that are resolved from their accompanying sources a fixed time $\Delta t$ after the peak of the microlensing event, which is also shown in Figure \ref{fig:mu_res}.
This, in turn, is fundamentally determined by what portion of the $\mu_{\rm rel}$ distribution the facility is able to sample, given its angular resolution.

There are two independent ways to obtain a relation that gives $M_{\ell}$ as a function of $D_{\ell}$.
First, $\theta_{\rm E}$ can be derived from a robust detection of finite-source effects from the observed microlensing light curve, which yields the angular size of the source star normalized to $\theta_{\rm E}$, and multiband photometry, from which the physical size of the source star can be determined.
Assuming the source is in the bar, $D_{s}$ is known to a precision equivalent to the width of the bar.
Then Equation (\ref{eq:mass_l}) simplifies to a mass-distance relation for the lens.
Secondly, a measurement of $F_{\ell}$ in conjunction with a mass-luminosity relation and an estimate of the extinction toward the lens provides another technique with which to compute the mass of the lens as a function of its distance.
Coupling these two methods uniquely determines $M_{\ell}$ and thus $M_{p}$.
Furthermore, measurements of both $D_{\ell}$ and $M_{\ell}$ together give the physical Einstein ring radius, which can then be used to determine $a_{\bot}$.

In the case of imaging a resolved lens, it is possible to directly measure the vector proper motion from the angular separation of the lens and source, the time elapsed since the peak of the event, and $u_{\rm 0}$.
When considering PSF elongation measurements and prompt follow-up photometry, I assume the magnitude of $\mu_{\rm rel}$ is known from Equation (\ref{eq:t_E}).
Then, in the case of the former, the elongation gives the flux ratio of the lens and source.
The source flux is measured from the microlensing light curve, although typically in a different bandpass than is used for the high-resolution photometry, thereby requiring an estimate of the source color.

\subsection{Imaging a Lens Spatially Resolved from the Source} \label{sec:imaging_resolved_lens}

\subsubsection{Practical Implementation of Technique}

A lens can be directly imaged after it is spatially resolved from the source.
The wait time $\Delta t$ after the closest approach of the source to the lens is at least several years for typical Galactic microlensing events.
This arises from the fact that it depends on both the relative proper motion of the two systems, which is generally $<$10 mas yr$^{-1}$ (see Figure \ref{fig:mu_res}), as well as the angular resolution attained by the observational facility, which is $\sim$100 mas for current facilities with the highest resolution.
In principle, after a resolved lens is imaged using a high-resolution facility, its measured apparent magnitude can be combined with a mass-luminosity relation and an estimate of the lens extinction to provide $M_{\ell}$ and, given an assumed $D_{s}$, $a_{\bot}$.

\subsubsection{My Approximated Methodology} \label{sec:irl_meth}

Here I take a lens to be resolved from the source when their angular separation satisfies
\begin{equation}
   \Delta\theta_{\ell,s} \equiv \sqrt{\left(\mu_{\rm rel}\Delta t\right)^{2} + \left(u_{\rm 0}\theta_{\rm E}\right)^{2}} \geq \theta_{\rm FWHM}.
\end{equation}
For all my simulated observing programs I assume that the minimum angular separation for the lens and source to be resolved is given by the $\theta_{\rm FWHM}$ of that facility, which is approximately equal to the $FWHM$ of an Airy Disc, given by $1.028\lambda/D$.

Next I determine the apparent $H$-band magnitude of the lens, $H_{\ell}$.
For the planetary host star of each lens system, I use the 1 Gyr isochrone of \citet{baraffe1998,baraffe2002} to obtain the absolute $H$-band lens magnitude, $M_{H,\ell}$, given its mass $M_{\ell}$ (see \S3.1.3 of H2014a).
I then convert the $I$-band extinction toward the lens, $A_{I,\ell}$ (see \S3.1.4 of H2014a), to the $H$-band lens extinction, $A_{H,\ell}$, using the relations of \citet{cardelli1989} and assuming $R_{V} = 2.5$ \citep{nataf2013}.
Finally, I compute $H_{\ell}$ from $M_{H,\ell}$, $A_{H,\ell}$, and $D_{\ell}$ (see \S3.1.2 of H2014a).

I then simulate an observing program for each lens system that would be resolved from its source for several values of $\Delta t$.
For each facility, I determine $t_{\rm exp}$ and $SNR$ as described in their respective sections in \S \ref{sec:high_res}.
Lastly, I compute the fractional precision to which $F_{\ell}$ can be measured in $H$-band, $\sigma_{H_{\ell}}$, using each facility after each $\Delta t$ interval.

\subsection{Elongation of the PSF of the Unresolved Microlensing Target} \label{sec:psf_elongation}

\subsubsection{Practical Implementation of Technique}

It is not necessary to wait until the lens and source are spatially resolved to constrain $F_{\ell}$.
As $\Delta t$ increases, the combined PSF of the unresolved lens and source will become distorted on a time scale dictated by $\mu_{\rm rel}$.
In the regime in which $\Delta\theta_{\ell,s} < \theta_{\rm FWHM}$, this elongation of the PSF of the microlensing target (lens+source) can be measured photometrically.
But, the PSF elongation itself stems from two factors: the separation of the lens and the source as well as their brightness ratio.
Thus, in order to constrain $F_{\ell}$ in this way it is necessary to obtain an independent measurement of one of these two causal parameters.
The lens-source separation can be determined by measuring $\mu_{\rm rel}$ as described in \S\ref{sec:lens_flux_methods}.
Then the elongation of the PSF subsequently gives the flux ratio of the lens and source.
Since the source magnitude is routinely derived from the ground-based light curve, $F_{\ell}$ can be computed (see \citealt{bennett2007} for a complete discussion).
Finally, as in the case of imaging a resolved lens, combining the inferred $F_{\ell}$ with a mass-luminosity relation and an estimate of the lens extinction yields $M_{\ell}$ and also $a_{\bot}$, assuming a source distance.

It is important to note that this technique hinges sensitively on the precision to which the morphology of the PSF is known.
Otherwise, any distortion of a PSF whose shape is poorly characterized could lead to a false-positive elongation measurement.
While I assume perfect knowledge of the PSF here, I concede that having sufficiently precise knowledge of the intrinsic PSF for a ground-based AO facility can prove extremely challenging.
This can be somewhat alleviated by the fact that typical bulge observing fields contain large samples of bright and isolated stars that can be used to model the PSF, but it may still be quite difficult to extensively model any spatial variations of the PSF.

\subsubsection{My Approximated Methodology} \label{sec:pe_meth}

In total there are four sources of uncertainty when using PSF elongation to constrain $F_{\ell}$:
\begin{enumerate}
   \item the statistical uncertainty of the source flux in the instrumental $I$-band, measured from the ground-based light curve,
   \item the uncertainty in calibrating the instrumental $I$-band source brightness,
   \item the uncertainty in transforming the $I$-band source brightness to the NIR filter of the high-resolution data, and
   \item the uncertainty on the fractional lens flux.
\end{enumerate}

The statistical uncertainty of the uncalibrated $I$-band magnitude of the source, determined from the modeling of the ground-based light curve, includes its covariances with other model parameters and is typically 2--5$\%$ (e.g., \citealt{dong2009,janczak2010,sumi2010,batista2011,yee2012}).
I take the typical fractional precision to be 2$\%$ to account for KMTNet's aperture size and higher cadence.
Then I take the sum of the uncertainty inherent to calibrating and transforming the uncalibrated ground-based $I$-band source brightness, items 2) and 3) from above, to be a conservative 0.03 mag \citep{janczak2010}.
I note that it is possible to improve on this in cases for which ground-based $H$-band data were taken during the event when the source was magnified, allowing $I-H$ to be computed to $\sim$1$\%$ \citep{batista2014}.

Defining the fractional lens flux as $f_{\ell} \equiv F_{\ell}/F_{\rm tot}$, where $F_{\rm tot} = F_{\ell} + F_{s}$ and $F_{s}$ is the flux of the source, the fractional precision of $f_{\ell}$ is given by
\begin{equation} \label{eq:sigma_lfrac}
   \sigma_{{f}_{\ell}} = \sqrt{\frac{2}{N_{\rm tot}}} \left(\frac{r_{\rm 0}}{\Delta\theta_{\ell,s}}\right)^{2} \frac{1}{\lvert1 - 2f_{\ell}\rvert},
\end{equation}
where $N_{\rm tot}$ is the total number of photons of the lens and source in the combined high-resolution PSF and $r_{\rm 0}$ is its Gaussian width \citep{bennett2007}.
This implies that $r_{\rm 0}$ is given by
\begin{equation}
   r_{\rm 0} = \frac{\theta_{\rm FWHM}}{2\sqrt{2{\rm ln}(2)}}.
\end{equation}

Computing both $N_{\rm tot}$ and $f_{\ell}$ requires $F_{s}$.
More specifically, it requires the apparent $H$-band magnitude of the source, $H_{s}$.
To determine $H_{s}$ for each microlensing event I first use the absolute $I$-band magnitude of the source $M_{I,s}$ (see \S3.1.1 of H2014a) and the same isochrone as in \S \ref{sec:irl_meth} to determine $M_{H,s}$, the absolute $H$-band magnitude of the source.
While rare, it is possible that $M_{I,s} < 2.67$, the bright limit of the isochrone, in which case I assume the source is a red clump giant.
I then use the absolute $I$-band and $H$-band magnitudes of the red clump, $M_{I} = -0.12$ \citep{nataf2013} and $M_{H} = -1.49$ \citep{laney2012}, to derive its intrinsic $I-H$ color, $I-H = 1.37$, from which I compute $M_{H,s}$.
$H_{s}$ and also $H_{\ell}$ are then determined from their respective absolute magnitudes using the procedure described in \S \ref{sec:irl_meth}.
I similarly compute $H_{\ell+s}$, the apparent magnitude of the lens and source combined in the single PSF, from which I obtain $N_{\rm tot}$.
Although both $f_{\ell}$ and $N_{\rm tot}$ could be affected by the contaminating flux of a blend star, for these computations I assume no such contribution.
In \S \ref{sec:contamination} I discuss the effect such a blend would have.

I subsequently simulate an observing program for each lens and source pair that would not be spatially resolved for several values of $\Delta t$.
The unresolved microlensing target is treated as a single point-source object whose brightness is the combined flux of the lens and the source, $F_{\ell+s} = F_{\ell} + F_{s}$.
I then use $H_{\ell+s}$ to determine $t_{\rm exp}$ and $SNR$ for the respective facilities as described in \S \ref{sec:high_res}.
Finally, I compute $\sigma_{H_{\ell}}$ by adding Equation (\ref{eq:sigma_lfrac}) in quadrature with the statistical uncertainty on $F_{s}$ (item 1) from above) and the uncertainty in calibrating and transforming $F_{s}$ (items 2) and 3) from above).

\subsection{Prompt High-resolution Follow-up Photometry} \label{sec:prompt_followup}

\subsubsection{Practical Implementation of Technique}

For cases in which the lens and source are unresolved and the PSF elongation is minimal, it is yet still possible to constrain $F_{\ell}$.
Both $F_{s}$ and $F_{\rm tot}$ are routinely measured from the ground-based microlensing light curve.
Then, a high-resolution image of the microlensing target will, to a high probability, resolve out all stars not dynamically associated with the microlensing event.
For reference, at the distance of the center of the Galactic bulge, $D_{\rm GC} = 8.2$ kpc \citep{nataf2013}, an angular separation of $\theta_{\rm FWHM,JWST}$ = 68 mas corresponds to a physical separation of 560 AU.
By assuming no companions to the lens or the source, any difference between the flux of the target measured in the high-resolution image and $F_{s}$ can be attributed solely to the lens.

In practice this requires taking high-resolution observations of the unresolved microlensing target after the peak of the event, typically in the NIR, and calibrating them.
The $I$-band flux of the source, which is routinely measured from the ground-based observed microlensing light curve data, must be transformed to the filter of the high-resolution data and also calibrated.
Then, the calibrated source flux can be subtracted from the high-resolution flux of the unresolved target, and any excess light can be attributed to the lens (see \S5 for a discussion of contaminating blend flux).
Each of these steps --- calibrating the high-resolution NIR data and transforming and calibrating the ground-based optical data --- introduces uncertainty that propagates through to the excess flux measurement.
A detection of the lens flux is taken to be secure only when the total uncertainty of the measured excess flux is small compared to the computed flux difference.
If $F_{\ell}$ is indeed robustly detected, $M_{\ell}$ and $a_{\bot}$ can be derived from a mass-luminosity relation and known values for the lens extinction and $D_{s}$.

\subsubsection{My Approximated Methodology}

A secure detection of $F_{\ell}$ via prompt follow-up photometry crucially requires careful treatment of the five sources of uncertainty involved in matching the ground-based and high-resolution data.
In addition to items 1)--3) discussed in \S \ref{sec:pe_meth} there is
\begin{enumerate}
   \setcounter{enumi}{3}
   \item the statistical uncertainty of the instrumental brightness of the unresolved microlensing target (lens+source) in the high-resolution data, and
   \item the uncertainty in calibrating the high-resolution measurement.
\end{enumerate}

As in \S \ref{sec:pe_meth}, I take the statistical uncertainty of the source flux to be 2$\%$ and the sum of the uncertainty inherent to calibrating and transforming the ground-based $I$-band source brightness to be 0.03 mag.
The final fractional precision of the calibrated $H$-band magnitude of the source, $\sigma_{H_{s}}$, is computed as the quadrature sum of these two uncertainties.

With regard to the high-resolution data, I compute the statistical uncertainty of the flux of the unresolved microlensing target via the methods described in \S \ref{sec:high_res}, wherein I assume the target to be a point source with flux equal to the combined flux of the lens and the source.
I conservatively assign a constant 0.03 mag uncertainty to the calibration process \citep{batista2011} and add the two in quadrature to obtain $\sigma_{H_{\ell+s}}$.
Finally, I define a lens flux detection via prompt follow-up photometry to occur when
\begin{equation} \label{eq:Delta_H}
   \Delta H \equiv H_{s} - H_{\ell + s} \geq N_{\rm sig,pfp} \times \sigma_{H,{\rm tot}},
\end{equation}
where $N_{\rm sig,pfp}$ represents the number of standard deviations at which the lens flux is detected and 
\begin{equation} \label{eq:sigma_H_tot}
   \sigma_{H,{\rm tot}} \equiv \sqrt{\sigma_{H_{s}}^{2} + \sigma_{H_{\ell + s}}^{2}}.
\end{equation}

\section{Results} \label{sec:results}

\subsection{Imaging a Lens that is Spatially Resolved from the Source} \label{sec:imaging_resolved_lens_results}

\begin{figure*}
   \centerline{
      \includegraphics[width=18cm]{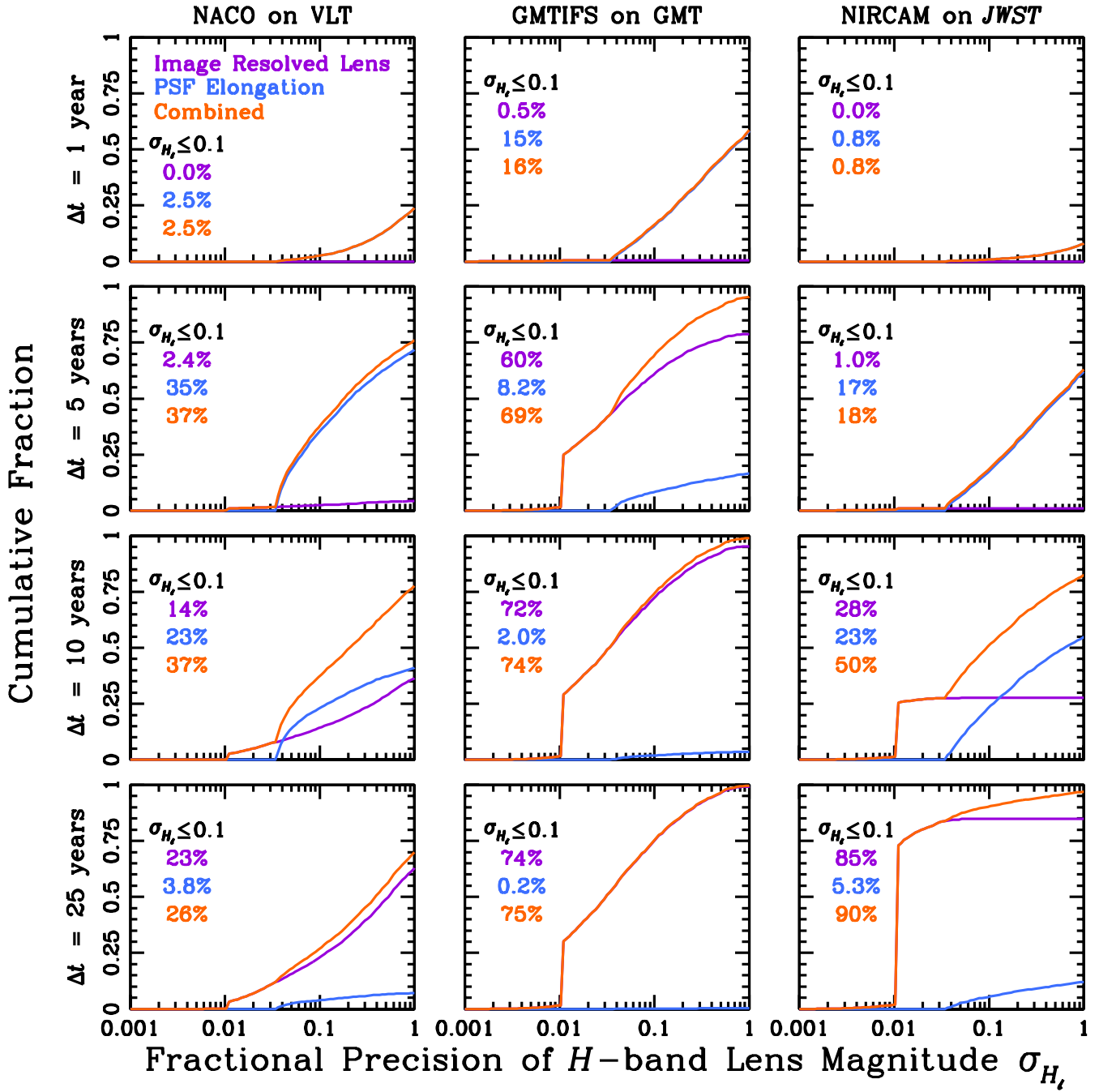}
   }
   \caption{
      \footnotesize{
         Cumulative distribution functions (CDFs) of $\sigma_{H_{\ell}}$ for imaging resolved lenses or measuring PSF elongation.
         The left-most column is for NACO on VLT, the middle shows GMTIFS on GMT, and the right-most NIRCAM on $JWST$.
         Each row represents a fixed time $\Delta t$ after the peak of the microlensing event.
         In each figure, the curves are color-coded according to technique.
         GMT will be able to image a majority of resolved lenses after $\Delta t = 5$ years and $JWST$ will be able to obtain $\sigma_{H_{\ell}} \leq 0.1$ for all lenses that are resolved after a given $\Delta t$ interval.
         Regarding PSF elongation, GMT can constrain $F_{\ell}$ to 10$\%$ or better for approximately one-seventh of predicted KMTNet detections after only $\Delta t$ = 1 year, and VLT and $JWST$ can do so for $\sim$35$\%$ and $\sim$17$\%$ of planetary systems, respectively, after $\Delta t$ = 5 years.
      }
   }
   \label{fig:irl_vs_pe}
\end{figure*}

Figure \ref{fig:irl_vs_pe} shows a cumulative distribution function (CDF) of $\sigma_{H_{\ell}}$ for observing programs that simulate imaging resolved lens systems $\Delta t$ = 1, 5, 10, and 25 years after the microlensing events using NACO on VLT, GMTIFS on GMT, and NIRCAM on $JWST$.
Although $JWST$ has the smallest aperture of the three facilities, its extremely low background allows it to achieve $\sigma_{H_{\ell}} \leq 0.1$ for all lenses that are resolved from their source after a fixed $\Delta t$.
Furthermore, the smaller background means the total exposure time required to do so is reduced compared to VLT and GMT.
For example, given the assumptions of my simulated observing programs and using the normalized planet detection rates computed by H2014a, after $\Delta t$ = 10 years it would take VLT $\sim$31 hours to image $\sim$25 planetary systems while it would take $JWST$ only $\sim$3.8 hours to image $\sim$18 planetary systems, and the majority of those imaged with VLT (about two-thirds) would have $\sigma_{H_{\ell}} > 0.1$.
The total fraction of events that can be observed after a fixed $\Delta t$ with VLT or $JWST$ rises as $\Delta t$ increases from 5 to 10 to 25 years, stemming from the fact that their values of $\theta_{\rm FWHM}$ are sampling the high proper motion tail ($\gtrsim$10 mas yr$^{-1}$), the peak ($\gtrsim$6), and the low proper motion tail ($\gtrsim$2), for those respective $\Delta t$ intervals.

GMT, on the other hand, will have have a collecting area $\sim$7.5 times bigger than that of VLT and $\sim$15 times bigger than that of $JWST$.
Additionally, if GMT will be able to achieve diffraction-limited imaging in $H$-band, it it will have a $\theta_{\rm FWHM}$ that is $\sim$3 and $\sim$4 times smaller than that of VLT and $JWST$, respectively.
Figure \ref{fig:irl_vs_pe} shows the result of the confluence of these two factors.
After $\Delta t$ = 5 years, GMT's diffraction limited resolution of $\theta_{\rm FWHM} = 16$ mas allows it to resolve all events with $\mu_{\rm rel} \gtrsim 3$ mas yr$^{-1}$, or $\sim$79$\%$ of the total planet detection rate.
GMT would be able to measure the flux of three-fourths of those events (or $\sim$60$\%$ of the total event rate) to a precision of $\sigma_{H_{\ell}} \leq 0.1$.
Again using the normalized planet detection rates of H2014a, after $\Delta t$ = 5 years GMT would be able to image the host star for $\sim$51 planetary systems whose lens is resolved from the source, $\sim$38 of those to a precision better than 10$\%$, and would be able to do so in $\sim$39 hours, given my assumptions for the simulated observing programs.
On the other hand, after $\Delta t$ = 5 years, VLT and $JWST$ could image only $\sim$3 and $\sim$2 total planetary systems with spatially resolved lenses, respectively.

Thus, for a fixed $\Delta t$, GMT will be able to obtain direct lens flux measurements for a significantly larger fraction of predicted KMTNet planet detections than VLT or $JWST$.
The primary benefit of VLT is that it exists and so could start observing lens systems shortly after KMTNet comes online, as early as the 2015 Galactic Bulge observing season, which begins in early February.
The advantage of $JWST$ rests in its ability to obtain $\sigma_{H_{\ell}} \leq 0.1$ for all lenses that are resolved after a given $\Delta t$, that its diffraction-limited capabilities do not hinge on favorable weather conditions or guide star characteristics, and the resulting shorter observing program required to image a fixed number of resolved lens systems.

\subsubsection{Physical Properties of Imaged Lens Systems with High-precision Flux Measurements}

\begin{figure*}
   \centerline{
      \includegraphics[width=19cm]{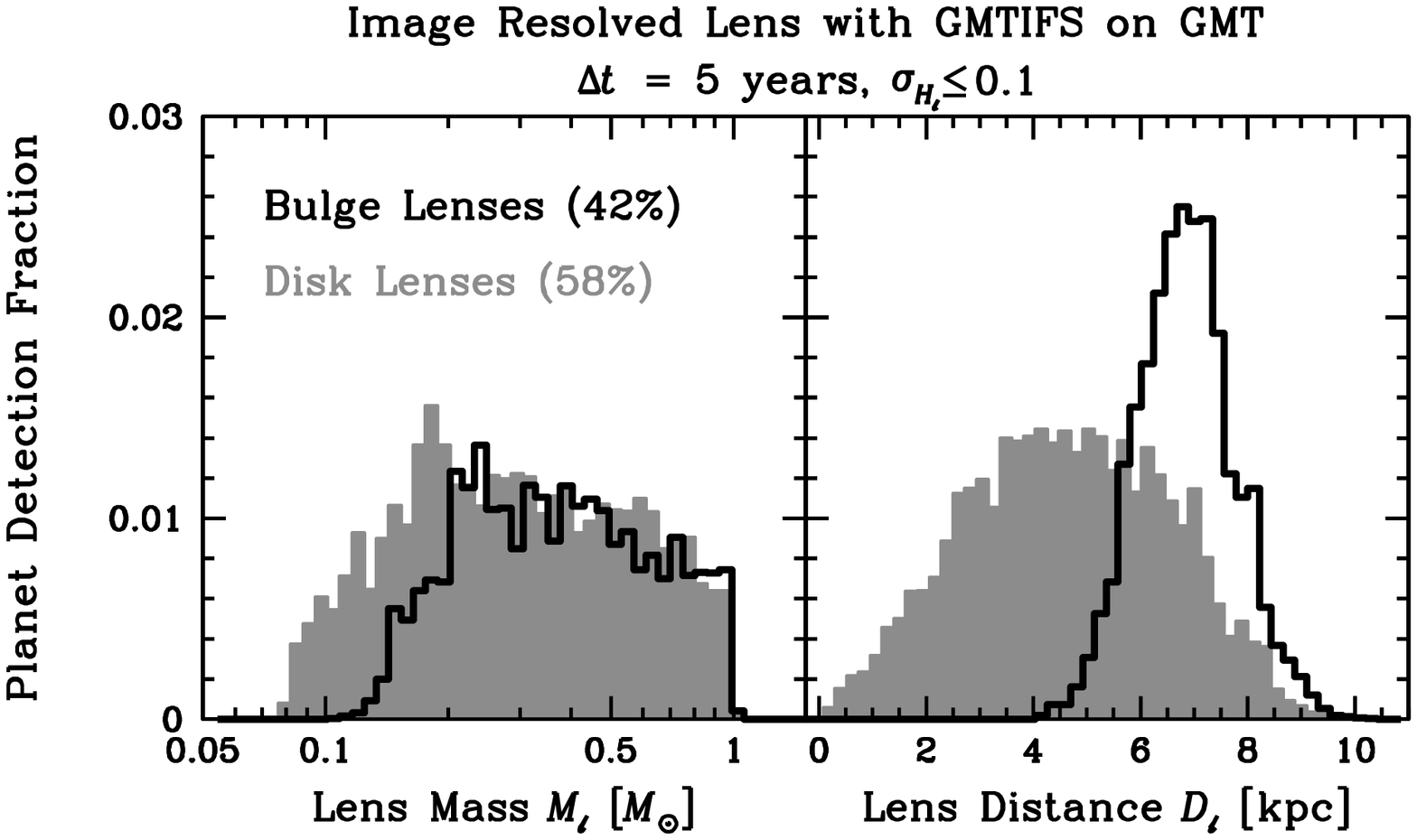}
   }
   \caption{
      \footnotesize{
         Distributions of $M_{\ell}$ and $D_{\ell}$ for resolved lens systems whose flux can be measured to a precision of $\sigma_{H_{\ell}} \leq 0.1$ using GMTIFS on GMT $\Delta t$ = 5 years after each event.
         The fraction of disk lenses (58$\%$) is substantially higher than the 45$\%$ of all predicted KMTNet planet detections that reside in the disk.
         This stems from the fact that closer lenses will be brighter for fixed $M_{\ell}$ and extinction, which is confirmed by the fact that the average lens distance is 0.5~kpc closer than the mean distance of 6.1~kpc for the full sample of predicted planetary systems.
         Furthermore, the disk lens population able to be imaged thusly will probe down to lower masses than will the bulge lens sample.
      }
   }
   \label{fig:irl_ig_mldl_rate}
\end{figure*}
I also examine the physical properties of the planet detections whose host star fluxes can be measured to a precision of $\sigma_{H_{\ell}} \leq 0.1$ by an example observing program.
Figure \ref{fig:irl_ig_mldl_rate} shows the distributions of $M_{\ell}$ and $D_{\ell}$ for spatially resolved lenses whose flux can be measured to $\leq$10$\%$ for my simulated observing program using GMTIFS on GMT after $\Delta t$ = 5 years.
The majority of such planetary systems that will be accessible by such an example observing program reside in the Galactic disk while the remaining 42$\%$ of lens systems will be bulge lenses.
In contrast, only 45$\%$ of the predicted KMTNet planet detections are expected to arise from disk lenses.
This increase in the fraction of disk lenses intuitively stems from the fact that disk lenses are generally closer, facilitating flux measurements that can be obtained with better precision.
This is corroborated by the distribution of $D_{\ell}$ for spatially resolved lenses versus that for the overall KMTNet planet detection sample.
The former has an average lens distance of $D_{\ell} = 5.6$~kpc while the latter has a mean distance of $D_{\ell} = 6.1$~kpc.
Thus, an observing program to image spatially resolved lenses will preferentially select for closer lens systems that are more likely to reside in the Galactic disk rather than the bulge.

Furthermore, the distributions of $M_{\ell}$ for disk and bulge lenses differ at the low-mass end.
The simulated observing program predicts that precise flux measurements will be possible for stars down to the Hydrogen burning limit at $\sim$0.08~$M_{\odot}$ for lenses in the Galactic disk.
However, the least massive bulge lenses able to be imaged thusly have masses that are $\sim$50$\%$ higher, around 0.13 $M_{\odot}$.
While this is to be expected, as more massive stars will be brighter, this indicates another implicit bias in the properties of lens systems whose masses will be derived from photometric flux measurements.
Those at larger distances that generally reside in the bulge will have, on average, higher masses than nearby disk lenses, for which it will be possible to probe planetary systems whose host stars have lower mass.
Understanding and accounting for these underlying selection effects when undertaking studies of the global properties of exoplanet detections will be of critical importance.

\subsection{Elongation of the PSF of the Unresolved Microlensing Target} \label{sec:psf_elongation_results}

Figure 2 also shows a CDF of $\sigma_{H_{\ell}}$ for observing programs that estimate measuring the PSF elongation of the unresolved microlensing target for $\Delta t$ = 1, 5, 10, and 25 years, again using NACO on VLT, GMTIFS on GMT, and NIRCAM on $JWST$.
For GMT, after $\Delta t$ = 25 years essentially all lenses and sources will be resolved given $\theta_{\rm FWHM,GMT}$, precluding PSF elongation measurements.
However, the first two terms of Equation (\ref{eq:sigma_lfrac}) contain the implicit scaling $\sigma_{f_{\ell}} \propto D^{-2.5}$, where $D$ is the diameter of the telescope aperture.
Thus, not only will the elongation of the PSF of the microlensing target be more pronounced (accounting for $D^{-2}$), but significantly more photons of the target will be collected for a fixed $t_{\rm exp}$ (yielding the remaining $D^{-1/2}$).
GMT is consequently able to measure $F_{\ell}$ to $\leq$10$\%$ for about one-seventh of planet detections merely one year after the event.
This presents a huge boon for future ground-based microlensing surveys, particularly KMTNet, and their ability to convert mass ratios $q$ to planet masses $M_{p}$ on expeditious time scales.

\subsection{Prompt High-resolution Follow-up Photometry} \label{sec:prompt_followup_results}

\begin{figure}
   \centerline{
      \includegraphics[width=9cm]{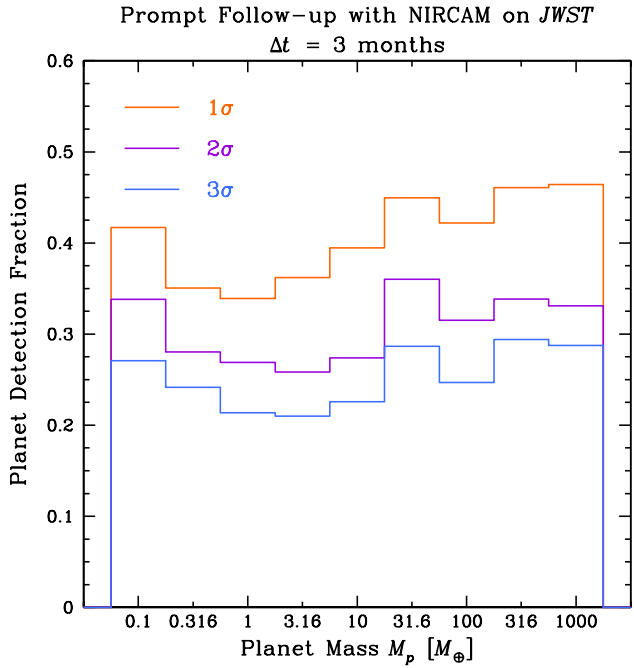}
   }
   \caption{
      \footnotesize{
         Fraction of planet detections in each $M_{p}$ bin for which $F_{\ell}$ can be constrained via prompt follow-up photometry taken after $\Delta t$ = 3 months with NIRCAM on $JWST$.
         The three different histograms correspond to three difference levels of significance for the lens flux detection, from Equations (\ref{eq:Delta_H}--\ref{eq:sigma_H_tot}).
         The results are nearly identical for NACO on VLT and GMTIFS on GMT, given that $\sigma_{H_{s}}$ and $\sigma_{H_{\ell + s}}$ are largely set by the uncertainties in calibrating and transforming the ground-based optical data and calibrating the high-resolution NIR data, respectively, rather than the photometric precision of each facility.
      }
   }
   \label{fig:pfp_ig}
\end{figure}

In Figure \ref{fig:pfp_ig} I show the fraction of lens systems for which it will be possible to securely detect the flux of the lens as a function of planet mass $M_{p}$ for three different $N_{\rm sig,pfp}$ detection thresholds using NIRCAM on $JWST$.
I consider lens flux detections with significances as low as $N_{\rm sig,pfp} = 1$ to explore cases in which an upper limit on the brightness of the lens can be established, even if the flux measurement itself is less secure.
For $N_{\rm sig,pfp} = 1$, indicating that $F_{\ell}$ is detected at the one-sigma level according to Equation (\ref{eq:Delta_H}), it will be possible to measure $F_{\ell}$ for $\sim$42$\%$ of planet detections in each mass bin.
This decreases to $\sim$31$\%$ for $N_{\rm sig,pfp} = 2$ and $\sim$26$\%$ for $N_{\rm sig,pfp} = 3$.
In all cases, though, it is approximately constant as a function of $M_{p}$, indicating no preference for or against certain planetary systems (as expected).
Furthermore, these fractions are essentially equivalent for NACO on VLT and GMTIFS on GMT, stemming from the fact that $\sigma_{H_{s}}$ and $\sigma_{H_{\ell + s}}$ are dominated by the uncertainties in calibrating and transforming the ground-based optical data and calibrating the high-resolution NIR data, respectively, rather than the $SNR$ each individual facility is able to achieve.

\section{Potential Sources and Effects of Contaminating Blend Light} \label{sec:contamination}

In the above scenarios I have ignored the possible contributions of additional flux from stars blended with the lens and/or source.
However, their presence could affect the measured fluxes and the masses ultimately derived from them.
Here I investigate the three most likely scenarios and the effect each would have.

\subsection{Lens Companion} \label{sec:lens_companion}

In principle, each lensing system could contain an additional stellar component whose flux could interfere with the derived value of $M_{\ell}$.
To test this, I begin by populating the lens system of each planet detection from H2014a with such a companion, all of whose parameters I will designate using the subscript ``$\ell_{2}$.''
Here I explore the impact for prompt follow-up photometry taken $\Delta t$ = 3 months after the time of the microlensing event.
Even with GMT, which has the smallest $\theta_{\rm FWHM}$ of the facilities I investigate, fewer than 0.01$\%$ of lens systems will be resolved from their source after 3 months.

\subsubsection{Implementation} \label{sec:lens_companion_method}

As described in H2014a and \S \ref{sec:kmtnet_galmod}, the lens masses are derived from the MF of \citet{gould2000}.
I draw the mass of the companion, $M_{\ell_{2}}$, from the same MF.
If the companion is a stellar remnant or BD, which I do not exclude as viable companions, I assume it to be dark and set its apparent magnitude accordingly, $H_{\ell_{2}} = 30$.
Otherwise, I determine $H_{\ell_{2}}$ via the procedure described in \S \ref{sec:irl_meth}.

Next I determine the parameters of the binary system comprised of the primary lens mass and this companion, which I designate as $\ell_{\rm bin}$.
I compute the orbital period $P_{\ell_{\rm bin}}$ from the log-normal Gaussian distribution of \citet{raghavan2010}, which has a mean of log $P$ = 5.03 and $\sigma_{{\rm log}~P} = 2.28$, where $P$ is in days.
From $P_{\ell_{\rm bin}}$ and $M_{\ell_{\rm bin}}$ I compute $a_{\ell_{\rm bin}}$.
I assume a circular orbit and compute $a_{\ell_{\rm bin},{\bot}}$ according to
\begin{equation}
   a_{\ell_{\rm bin},{\bot}} = a_{\ell_{\rm bin}} \sqrt{1 - {\rm cos}^{2}\zeta}.
\end{equation}
For randomly oriented orbits, ${\rm cos}\zeta$ is uniformly distributed, so I therefore draw ${\rm cos}\zeta$ from a uniform random deviate in the range [0--1].
The mass ratio $q_{\ell_{\rm bin}}$ is simply $M_{\ell_{2}}/M_{\ell}$.

\subsubsection{Occurrence Probability and Effect on Derived Lens Mass} \label{sec:lens_companion_effect}

In many cases it will be possible to detect the presence of a lens companion aside from its flux contribution, circumventing errors introduced by an unseen companion.
The two primary ways this can be achieved are if the lens companion is spatially resolved from the unresolved microlensing target or if the source trajectory passes near enough to the central caustic that the perturbations to the caustic induced by the presence of this additional lensing mass are then observed in the light curve.
To determine the former I compute the angular separation of the lens companion and the unresolved microlensing target, $\Delta\theta_{\ell+s,\ell_{2}}$, after $\Delta t$ = 3 months and presume the companion would be detected if $\Delta\theta_{\ell+s,\ell_{2}} \geq \theta_{\rm FWHM}$.

Regarding the latter, I assume the lens companion would be detected if the source passes over or very near the central caustic perturbation induced by the companion's presence, specifically if $u_{\rm 0} \leq u_{{\rm cc},\ell_{2}}$.
For a two-body lens system there exists a set of 1--3 closed caustic curves, depending on the angular separation of the lensing masses, that identify the locations in the plane of the source where the magnification of a point-like source diverges to infinity.
There is one central caustic that is located near the center of mass of the lens system and 1--2 planetary caustics.
I first compute the topology of the $\ell_{2}$ binary \citep{erdl1993}, which determines the total number of caustics.
For all topologies I take $u_{{\rm cc},\ell_{2}}$ to be the size of the central caustic along its longest dimension.
In the case of a resonant topology, for which the central caustic is the sole caustic, I compute $u_{{\rm cc},\ell_{2}}$ numerically.
If the topology is close or wide, I find the approximate dimensions of the central caustic analytically using Equations (22--23) or (9--10) of \citet{bozza2000}, respectively.\footnote{The third term in Equations (9--10) of their manuscript contains $\rho_{i}^{2}$, where $\rho$ is their nomenclature for projected separation, when they should instead read $\rho_{i}^{3}$ as a result of their perturbative analysis. I have corrected this prior to my implementation of said analytic approximations.}

\begin{figure*}
   \centerline{
      \includegraphics[width=19cm]{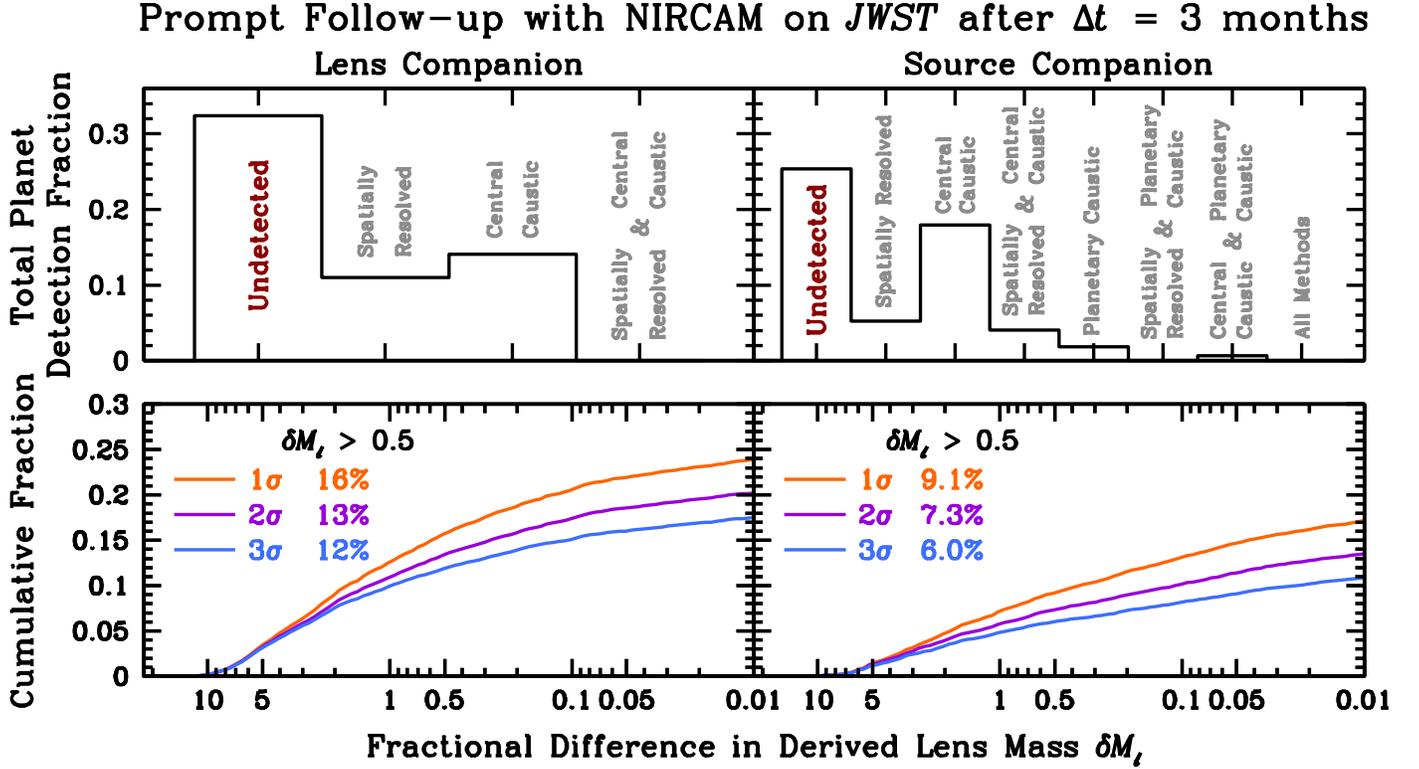}
   }
   \caption{
      \footnotesize{
         Detectability of lens (top left) and source (top right) companions and their respective resulting $\delta M_{\ell}$ CDFs (bottom).
         Assuming every lens or source has a companion, the companions to approximately half of the events with robust non-source flux detections will go undetected.
         However, the fraction of catastrophic failures in the derived values of $M_{\ell}$ is low, not exceeding (16$\cdot f_{\rm bin})\%$ of all planet detections, where $f_{\rm bin}$ is the binary fraction.
         This is then furthermore suppressed by empirically determined binary fractions, which are a steep function of spectral type and can be as low as $f_{\rm bin} \sim 25\%$ for M stars \citep{lada2006}.
      }
   }
   \label{fig:companions_nj}
\end{figure*}

I show the fraction of lens companions whose presence would be detected via the above methods in Figure \ref{fig:companions_nj}.
Here I have also excluded events whose lens and source would be resolved after $\Delta t$ = 3 months with $JWST$ (as a conservative estimate) as well as those for which the lens flux would not be detected at the one-sigma level or better (see \S \ref{sec:prompt_followup}), again via $JWST$ (though the choice of facility doesn't affect this criterion, as discussed in \S \ref{sec:prompt_followup_results}).
This leaves $\sim$57$\%$ of planet detections, which is roughly one-third higher than what is presented in Figure \ref{fig:pfp_ig}.
The inclusion of a companion to the lens increases the flux of the non-source term in Equation (\ref{eq:Delta_H}), in turn increasing $\Delta H$ and, consequently, the overall fraction of systems for which non-source flux would be robustly detected.
For those cases in which the lens companion would be spatially resolved from the unresolved microlensing target, I have not included its flux contribution when determining if non-source flux is robustly detected.
Of this $\sim$57$\%$, lens companions for about half of these events, or $\sim$32$\%$ of all planet detections, will go undetected, with comparable fractions of companions being spatially resolved or detected by the source passing sufficiently near the portion of the central caustic for which the presence of the companion will manifest itself in the light curve.

For the $\sim$32$\%$ of lens systems in which $\ell_{2}$ is undetected, I estimate the fractional uncertainty the blend flux from this undetected companion introduces to the derived lens mass.
I compute $M_{H_{\ell + \ell_{2}}}$, the absolute magnitude of the combination of the lens and its companion, from $D_{\ell}$, $A_{H_{\ell}}$, and $H_{\ell + \ell_{2}}$.
Using the same isochrone as in \S \ref{sec:irl_meth} I determine $M_{\ell,{\rm blend}}$, the mass of the primary lens that would be inferred if the blend flux contributed by the companion were undetected and otherwise attributed to the lens.
I then compute the absolute fractional difference between the true primary lens mass and the mass determined when including blend flux from the undetected lens companion,
\begin{equation}
   \delta M_{\ell} = \frac{\lvert M_{\ell} - M_{\ell,{\rm blend}} \rvert}{M_{\ell}}.
\end{equation}
Figure \ref{fig:companions_nj} shows the resulting CDF of $\delta M_{\ell}$ for three different values of $N_{\rm sig,pfp}$.

The CDF has been truncated at $\delta M_{\ell} = 0.01$, which explains why the right-most limit is lower than the height of the ``Undetected'' bin, for two reasons.
First, this threshold is comparable to the finest steps in stellar mass of the isochrone.
Secondly, of the 32 planets that have hitherto been detected via microlensing,\footnote{From \url{http://exoplanet.eu} as of 8/September/2014} all mass values have fractional uncertainties greater than 4$\%$, indicating that a $\sim$1$\%$ uncertainty is an appropriate floor for the precision with which it is currently possible to obtain $M_{p}$ for microlensing planet detections.
I take $\delta M_{\ell} > 0.5$ to define lens systems for which there is a catastrophic failure in the determination of $M_{\ell}$ due to contaminating flux from an undetected companion to the lens.
After excluding systems whose lens and source are resolved, whose non-source light is not detected at one sigma or better, and lens companions whose presence would be otherwise noticed, either by being spatially resolved or via perturbing the central caustic, I find $\lesssim$16$\%$ of lens systems will have their masses severely mis-estimated with $\delta M_{\ell} > 0.5$.

\begin{figure}
   \centerline{
      \includegraphics[width=9cm]{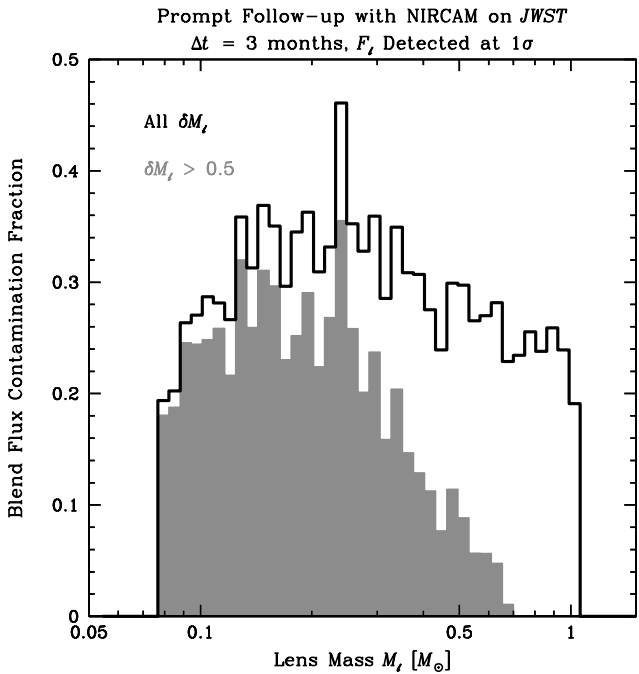}
   }
   \caption{
      \footnotesize{
         Fraction of lenses subjected to contamination from the blend flux of undetected lens companions as a function of $M_{\ell}$.
         For derived lens masses that suffer from catastrophic failures, with $\delta M_{\ell} > 0.5$, there is a steep dependence on $M_{\ell}$.
      }
   }
   \label{fig:klp_lc_nj_contaminatedfrac_ml}
\end{figure}
Figure \ref{fig:klp_lc_nj_contaminatedfrac_ml} shows the fraction of lens systems with undetected lens companions as a function of the true lens mass.
On average, $\sim$30$\%$ of lenses with a given mass will be potentially subjected to flux contamination from a lens companion.
There is a strong dependence on $M_{\ell}$ when considering only systems whose mass derivations are subject to catastrophic failures from these undetected companions.
The bulk of detections with $\delta M_{\ell} > 0.5$ have $M_{\ell} \lesssim 0.3M_{\odot}$ and none have $M_{\ell} \gtrsim 0.7M_{\odot}$.
Lenses with the lowest masses will have more massive companions that are thus more luminous, leading to a higher probability of significant contamination from the blend flux.
Conversely, lenses near the high-mass end of the MF will generally have lower-mass companions.
The mass-luminosity relation is sufficiently steep that the light from the lower-mass companions to these massive lenses will not significantly skew the derived lens mass.

For this calculation I have assumed that each primary lens star has exactly one companion (in addition to the planet).
However, not only is the binary fraction, $f_{\rm bin}$, $<$100$\%$, but it also depends steeply on spectral type (e.g., \citealt{duquennoy1991,fischer1992,lada2006,raghavan2010} and references therein).
In fact, M stars, which comprise the bulk of the Galactic lens population, can have a binary fraction as low as $\sim$25$\%$ \citep{lada2006}.
While Figure \ref{fig:klp_lc_nj_contaminatedfrac_ml} shows contamination fraction as a function of $M_{\ell}$, I allow for lens companions that are more massive than the planet host star.
In these cases the host star would not be the primary star of the stellar binary.
Thus, although the higher fraction of catastrophic lens mass derivation failures for low-mass lenses is caused by brighter, higher-mass lens companions, it is these companions --- {\it not} the lens host stars themselves --- that are the primary bodies in the stellar binaries, and as such they will have different values of $f_{\rm bin}$.
Digesting any potential bias in the distribution of photometrically derived lens masses is thus quite complicated and will likely require a global approach rather than being addressed system-by-system.

Whatever the value of $f_{\rm bin}$, it is crucial to also note that the CDF in Figure \ref{fig:companions_nj} will be suppressed by that same factor.
This establishes the results presented here as an upper limit.
Therefore, while individual systems may yet experience catastrophic failures in their mass determinations, this indicates that undetected lens companions will have a small net effect on derived values of $M_{\ell}$ for the statistically large samples of planet detections H2014a predict KMTNet will find.

\subsection{Source Companion} \label{sec:source_companion}

Another possibility hitherto unaccounted for is the contribution of additional flux from a companion to the source star.
As with a lens companion, if the presence of such a star goes undetected, its flux will skew the derived value of $M_{\ell}$.
Here I populate the source star of each detected planetary microlensing event predicted for KMTNet from H2014a with a companion, designating all of its parameters with the subscript $s_{2}$.
I again investigate the resulting effect only for prompt follow-up photometry $\Delta t = 3$ months after the time of each microlensing event.

\subsubsection{Implementation} \label{sec:source_companion_method}

I determine the mass of the source companion, $M_{s_{2}}$, and its apparent $H$-band magnitude, $H_{s_{2}}$, following the prescription described in \S \ref{sec:lens_companion_method}.
However, prior to obtaining the parameters of the binary source $s_{\rm bin}$, I must determine the mass of the source star itself.
I draw its absolute $I$-band magnitude $M_{I,s}$ from the LF of \citet{holtzman1998} (see \S3.1.1 of H2014a), which I use in conjunction with the same isochrone as previously to obtain $M_{s}$.
If $M_{I,s} < 2.67$, the bright end of the isochrone, $M_{s_{2}}$ is taken to be 1.1$M_{\odot}$, typical for G and K giants.
With the masses of both components of $s_{2}$ in hand, I compute the binary parameters $P_{s_{\rm bin}}$, $a_{s_{\rm bin}}$, $a_{s_{\rm bin},{\bot}}$, and $q_{s_{\rm bin}}$ as laid out in \S \ref{sec:lens_companion_method}.

\subsubsection{Occurrence Probability and Effect on Derived Lens Mass} \label{sec:source_companion_effect}

I investigate three channels through which a companion to the source can be detected.
As with a lens companion, the simplest is if the source companion and the microlensing target are spatially resolved.
If $\Delta\theta_{\ell+s,s_{2}} \geq \theta_{\rm FWHM}$ after $\Delta t$ = 3 months I assume the companion to be detected.

Otherwise, I assume that the source companion would be detected if the source passes over the central caustic created by the lens host star and planet.
In this regime $q \ll 1$, so I make use of analytic approximations for the size of the central caustic, depending on the topology.
If it is a resonant topology, I compute $u_{{\rm cc},\ell}$ numerically.
Otherwise, if it is a close or wide topology, I use Equations (10--11) of \citet{chung2005} (or, equivalently, Equations (24--25) of \citealt{han2006}).
If $u_{\rm 0} \leq u_{{\rm cc},\ell}$, I assume the source companion would be detected via additional features in the light curve.

In many cases, however, the trajectory of the source will cause it to maintain a wide separation from the lensing star throughout the duration of the event (see Figure 23 of H2014a).
The detection of the planet then arises from the source passing near or over (at least) one of the planetary caustics.
Just as the presence of a source companion would manifest itself through extra magnification structure in the light curve as it passes over the central caustic, so would it if it were to pass over a planetary caustic, if $a_{s_{\rm bin},{\bot}}$ were sufficiently small.
There are no planetary caustics for a resonant topology.
For a close topology I approximate the size of the planetary caustic $u_{{\rm pc},\ell}$ using Equations (3), (15), and (18) of \citet{han2006}, noting that the caustic width along the axis parallel to the planet-star separation vector is always larger than the width along the perpendicular direction, obviating computation of the latter.
If it is a wide topology I compute $u_{{\rm pc},\ell}$ from Equation (8) of \citet{han2006}.
If the angular separation of $s_{2}$ normalized to $\theta_{\rm E}$ is smaller than the size of the planetary caustic, $\Delta\theta_{\ell+s,s_{2}}/\theta_{\rm E} \leq u_{{\rm pc},\ell}$, I assume that the source companion would induce detectable perturbations on the light curve.

In principle it is also possible to detect the presence of a source companion due to a shift in the observed color of the microlensing event.
Gravitational microlensing is itself achromatic.
However, in the case of a binary source it is likely that the flux ratio of the two components will deviate from one, indicating a difference in color between the two stars.
Then, a microlensing target whose total observed flux is color-dependent evinces the binarity of the source, which is more readily detectable as the binary source passes over or near the caustics.
While this effect has previously been measured \citep{hwang2013}, I do not consider it here and instead mention it as an additional tool with which source companions can be detected.

Figure \ref{fig:companions_nj} shows a histogram of the fraction of events whose companions to the source would be detected by combinations of the methods described above.
Similar to the consideration for lens companions, I include only events that remain unresolved after $\Delta t$ = 3 months and for which the lens flux would be detected at the one sigma level or better via prompt follow-up photometry using NIRCAM on $JWST$.
This represents a more conservative approach, given that the smaller values of $\theta_{\rm FWHM}$ for VLT and GMT cause them to spatially resolve more companions from sources.
I again note that the sum of the histogram bins gives a fraction that is larger than the one-sigma fraction shown in Figure \ref{fig:pfp_ig}, arising from the increase in $\Delta H$ (Equation (\ref{eq:Delta_H})) that is due to the increase in brightness of the non-source term.
Approximately 25$\%$ of source companions would go undetected, smaller than the $\sim$32$\%$ of lens companions.

If the source companion is indeed undetected, then its contributed blend flux will influence the derived value of $M_{\ell}$.
I then follow the same procedure as in \S \ref{sec:lens_companion_effect} to compute $\delta M_{\ell}$.
The resulting CDF is shown in Figure \ref{fig:companions_nj}, again truncated at $\delta M_{\ell} = 0.01$.
Even fewer undetected source companions would induce catastrophic failures in the eventual lens mass determination, with $\lesssim$9$\%$ of detections having $\delta M_{\ell} > 0.5$.
Additionally, the same caveat regarding the binary fraction $f_{\rm bin}$ applied to lens companions holds here, which would only further reduce the fraction of planetary systems for which prompt follow-up photometry would ultimately produce values of $M_{\ell}$ that would be catastrophically skewed.
However, it is important to note that the source stars of microlensing events have spectral types that are, in general, earlier than those of lens stars, leading to different binary fractions between the two populations.

\subsection{Ambient Interloping Star} \label{sec:ambient_interloper}

I lastly investigate the probability that a star not dynamically associated with the microlensing event could be blended with the microlensing target, even in a high-resolution image.
This has previously been estimated on a case-by-case basis for individual planet detections \citep{dong2009,sumi2010,janczak2010,batista2011}.
The approach taken is to count the number of stars on the high-resolution image within, e.g., 3$\sigma$ of the detected excess flux and estimate the probability that there could be one within the PSF of the microlensing target.
While the probability of such an occurrence has been $\lesssim$5$\%$ in all cases, a blend contribution from an ambient interloping star could be more insidious.
Rather than the possibility that $all$ of the excess flux could be due to an interloper, there exists the possibility that only some of it is.
Depending on the magnitude of the contribution, this could affect the derived lens mass in the same way as an undetected companion to the lens or source.

I begin by appending the LF of \citet{zheng2004} to that of \citet{holtzman1998}, normalizing the former to the latter using data in the range $6.5 \leq M_{I} \leq 9.0$, where the two overlap.
Then I convert the combined LF, which now extends to $M_{I} = 13.5$, to $H$-band using the same isochrone as in \S \ref{sec:irl_meth}, again using $I-H$ = 1.37 for stars with $M_{I} < 2.67$.
I take $A_{I} = 2.0$ to be typical of the proposed KMTNet fields (see Figure 13 of H2014a) and convert to $H$-band using the \citet{cardelli1989} relations and $R_{V} = 2.5$ \citep{nataf2013} to obtain $A_{H} = 0.8$.
Assuming a uniform distance to all interloping stars equivalent to the Galactocentric distance, $D_{\rm int} = R_{\rm GC} = 8.2$ kpc \citep{nataf2013}, I then compute the CDF for $\theta_{\rm FWHM,JWST}$, as it has the smallest aperture and thus largest $\theta_{\rm FWHM}$ of the facilities explored here.
I also multiply the stellar number density of the LF by a factor of 1.25 to account for the average increase in the surface density of stars of the KMTNet fields compared to that of Baade's Window, for which the LF was derived, using the Galactic density models described in \S3.1.2 of H2014a.

\subsubsection{Results} \label{sec:ambient_interloper_results}

\begin{figure}
   \centerline{
      \includegraphics[width=9cm]{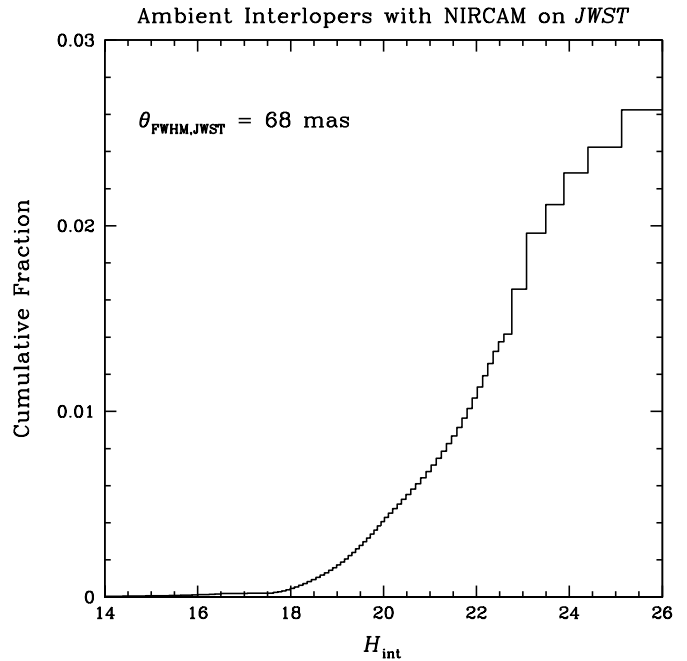}
   }
   \caption{
      \footnotesize{
         CDF of chance alignment with an ambient interloping star of magnitude $H_{\rm int}$.
         A pessimistic approximation of $\theta_{\rm FWHM}$ = 200 mas increases the maximum probability to $\sim$23$\%$, while a best-case scenario of $\theta_{\rm FWHM,GMT}$ = 16 mas reduces it to $\sim$0.15$\%$.
      }
   }
   \label{fig:interloper}
\end{figure}

Figure \ref{fig:interloper} shows a CDF of the probability of the presence of an ambient interloping blend star as a function of $H$-band magnitude of the interloper $H_{\rm int}$ as computed above.
The apparent magnitude distribution extends to $H_{\rm int} \sim 26$, which is equivalent to the faint limit of lenses that could be detected by any of the methods discussed here.
The probability of such a star falling within a seeing disc with FWHM = $\theta_{\rm FWHM,JWST}$ is $\lesssim$3$\%$ across the full magnitude range.
Decreasing the assumed values of $A_{H}$ and $D_{\rm int}$ only acts to shift the distribution to brighter magnitudes, leaving the maximum probability unaffected.
In assuming $\theta_{\rm FWHM} = 200$ mas as a worst-case scenario approximation, the CDF reaches a maximum probability of $\sim$23$\%$.
While this is undoubtedly more significant, it is still low and otherwise improbable, and even for smaller values of $D_{\rm int}$ and $A_{H,{\rm int}}$ the net effect, i.e., the resulting $\delta M_{\ell}$, is likely non-catastrophic.
More optimistically, using $\theta_{\rm FWHM,GMT}$ yields a maximum probability of $\sim$0.15$\%$.

\section{Discussion} \label{sec:discussion}

Here I have explored the potential of current and future high-resolution facilities to obtain flux measurements of the host stars of planetary systems predicted to be detected by KMTNet (H2014a).
GMTIFS on GMT provides a powerful tool with which to constrain lens fluxes.
It will be able to measure $F_{\ell}$ to $\leq$10$\%$ for $\sim$60$\%$ of KMTNet's predicted planet detections $\Delta t$ = 5 years after each event by imaging lenses spatially resolved from the source, and for roughly one-seventh of detections after $\Delta t$ = 1 year by measruing the elongation of the PSF of the unresolved microlensing target (lens+source).
Furthermore, NIRCAM on $JWST$ would be able to carry out high-precision ($\sigma_{H_{\ell}} \leq 0.1$) measurements for $\sim$28$\%$ of events $\Delta t$ = 10 years after each event by imaging resolved lenses, and NACO on VLT could obtain lens flux measurements via prompt follow-up photometry for the $\sim$42$\%$ of planet detections accessible to it at the one-sigma level within $\Delta t$ = 3 months of the events and could be used as soon as KMTNet comes online.
These are exciting prospects for increasing the number of well-constrained microlensing planet detections, which themselves are integral to our understanding of their underlying demographics and formation mechanisms.

I additionally explore the effects contaminating flux from possible blended objects would have on $F_{\ell}$.
Undetected companions to the lens would lead to catastrophic failures in the derived lens mass, $\delta M_{\ell} > 0.5$, for $\lesssim$(16$\cdot f_{\rm bin})\%$ of predicted KMTNet planet detections.
The same fraction for undetected companions to the source drops to $\lesssim$(9$\cdot f_{\rm bin})\%$.
In both cases I have assumed 100$\%$ binarity, so these fractions would be further suppressed by the underlying distribution of stellar multiplicity.
The integrated probability of blend flux contributions from interloping stars not dynamically associated with the event is even lower, reaching a maximum of $\sim$3$\%$ for $\theta_{\rm FWHM,JWST}$ = 68 mas.

\subsection{Measuring Lens Masses with Parallax and Proper Motion}

In this paper I have focused on methods to constrain lens fluxes.
By combining measurements of $F_{\ell}$ and $\theta_{\rm E}$ with a mass-luminosity relationship and an estimate of the extinction toward the lens, $M_{\ell}$ can be derived via Equation (\ref{eq:mass_l}).
It is possible to also obtain an independent measurement of the lens mass from information obtained when imaging a resolved lens.
Because the lens and source are resolved, their angular separation $\Delta\theta$ can be computed from the photometric images.
The time elapsed since the peak of the microlensing event $\Delta t$ is also known.
From these two parameters the vector heliocentric proper motion can be measured.
The direction of proper motion is parallel to the parallax vector.
So, a measurement of the vector proper motion combined with a one-dimensional measurement of the component of the microlens parallax that is parallel to the direction of the Earth's acceleration, $\pi_{\rm E,\parallel}$, yields a direct measurement of $M_{\ell}$ \citep{gould2014a}.

There are several advantages to this method.
It does not rely on a detection of finite-source effects, it does not require multiband data, it is not subject to the systematic uncertainties inherent in the conversion of a source color to a physical radius, and the presence of a companion to the lens and/or source does not introduce additional uncertainties (in fact, it is precisely the opposite if said companions are bright).

Here I explore the ability of a simulated observing program on GMT $\Delta t$ = 5 years after each event to compute the vector proper motion and $\pi_{\rm E,\parallel}$, which ultimately give a direct measurement of $M_{\ell}$.
There are three quantities involved in this process: 1) the magnitude of the proper motion, 2) the direction of the proper motion, and 3) the one-dimensional parallax.
Given that $\Delta t$ will be known to extremely high precision, the two primary sources of uncertainty will be from the vector proper motion and $\pi_{\rm E,\parallel}$.
As shown in \citet{gould2003}, the asymmetry induced by parallax can be encapsulated in a single parameter $\gamma$ that is proportional to $\pi_{\rm E,\parallel}$.
They provide an analytic scaling relation\footnote{Their manuscript indicates that $\sigma_{\gamma} \propto f$ when it should read $\sigma_{\gamma} \propto f^{-1}$, i.e., a higher observational cadence acts to improve the precision to which $\gamma$ can be measured. I have corrected this in the above equation.} to estimate the fractional precision to which $\gamma$ can be determined from a dedicated ground-based microlensing observational campaign,
\begin{IEEEeqnarray}{rCL} \label{eq:sigma_gamma}
   \frac{\sigma_{\gamma}}{\lvert\gamma\rvert} & = & \frac{1}{12} \left(\frac{\sigma_{\rm ph}}{0.01}\right) \left(\frac{f}{144~{\rm day}^{-1}}\right)^{-1} \left(\frac{S}{3}\right) \left(\frac{\tilde{v}}{800~{\rm km}~{\rm s}^{-1}}\right) \nonumber\\ && \times\> \left(\frac{t_{\rm E}}{20~{\rm days}}\right)^{-3/2} \left(\frac{\lvert{\rm cos}~\psi~{\rm cos}~\phi\rvert}{0.5}\right)^{-1},
\end{IEEEeqnarray}
where $\sigma_{\rm ph}$ is the photometric precision of the target, $f$ is the cadence of observations, $S$ will vary monotonically between 2.1 and 4.4 for typical KMTNet observations, $\tilde{v}$ is the relative lens-source velocity projected onto the observer plane, ${\rm cos}~\psi$ gives the length of the Earth-Sun separation projected onto the plane of the sky, and $\phi$ is the angle between the source trajectory and said projected separation.
I determine $\sigma_{\rm ph}$ for each event as described in \S3.3.2 of H2014a (excluding noise due to the Moon and to unresolved stars) and estimate $f$ to be 54 day$^{-1}$, assuming an average nine-hour observing night and a ten-minute cadence for KMTNet's three telescopes.
I take $S$ to be 3 and compute $\tilde{v}$ for each event via
\begin{equation}
   \tilde{v} = \frac{\theta_{\rm E}D_{\ell}}{t_{\rm E}} \left(\frac{D_{s}}{D_{s} - D_{\ell}}\right).
\end{equation}
Lastly, I set the final term equal to the fiducial value of 0.5 for each event.

I assume that the fractional precision of the magnitude of $\mu_{\rm rel}$ is the quadrature sum of the precisions to which the centroids can be determined for both the lens and the source, which I approximate as the ratio of the FWHM (in pixels) to the $SNR$, divided by $\Delta t$.
The uncertainty in centroiding both the lens and the source is included twice, once for each component axis.
The fractional precision of the direction of $\mu_{\rm rel}$ is similar to the fractional precision of its magnitude, so I multiply the latter by $\sqrt{2}$ to obtain the fractional precision of the vector proper motion.
Because the vector proper motion measurement comes from the high-resolution data, the $SNR$ for both the lens and source is computed as described in \S \ref{sec:nextgen_ground}.

\begin{figure*}
   \centerline{
      \includegraphics[width=19cm]{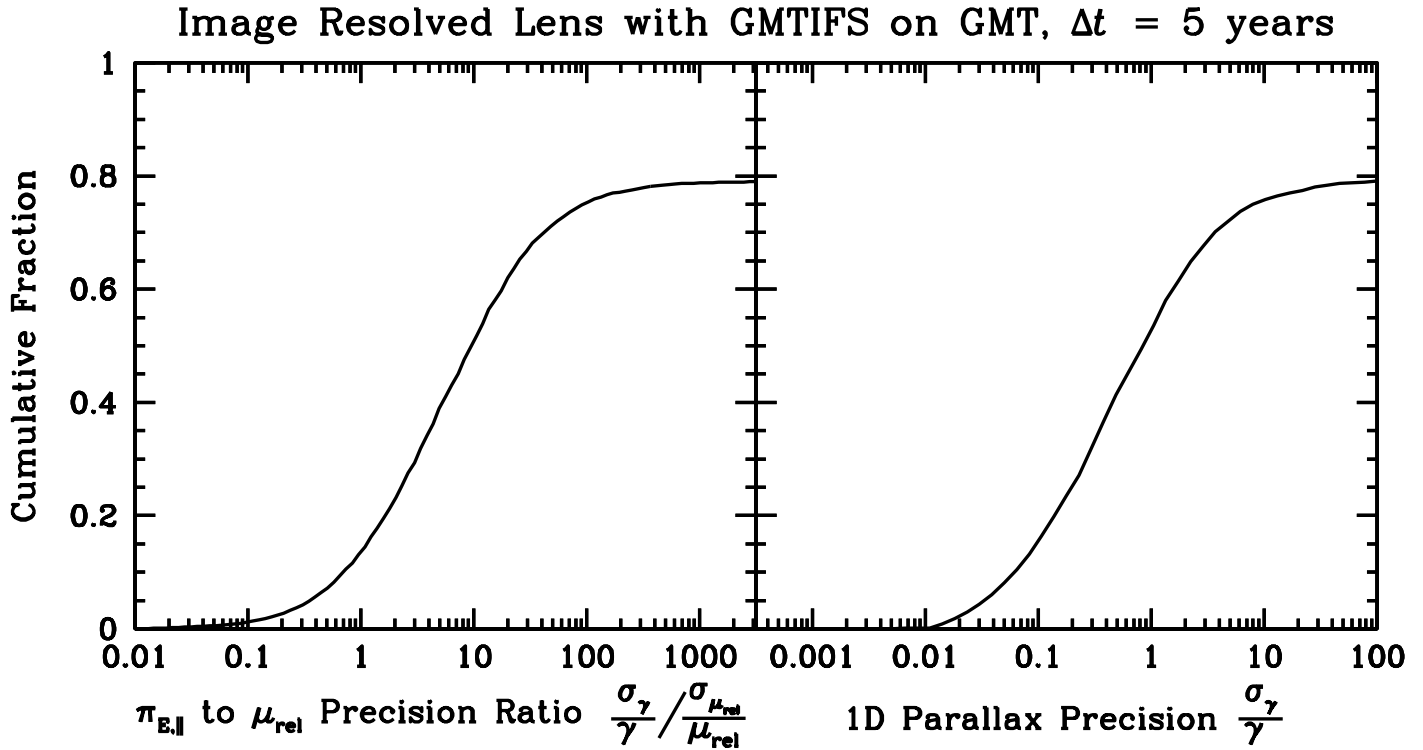}
   }
   \caption{
      \footnotesize{
         CDF of the ratio of the fractional precision of the one-dimensional parallax to that of the vector proper motion (left) and of the fractional precision of $\gamma$ alone (right).
         In general the one-dimension parallax $\pi_{\rm E,\parallel}$ is known to poorer precision than is $\mu_{\rm rel}$.
         Only $\sim$14$\%$ of events would have lens masses known to $\leq$10$\%$ using this method, under my assumptions.
      }
   }
   \label{fig:sigma_ratio_gamma_cdf}
\end{figure*}

In practice, the fractional precision of a lens mass derived in this manner requires careful treatment of the covariances between all input parameters.
Furthermore, the measured proper motion vector is derived in a heliocentric reference frame whereas $\pi_{\rm E,\parallel}$ and $\Delta t$ are in a geocentric frame.
Transforming between the two requires solving a quadratic equation and can thus lead to a potential two-fold degeneracy in certain cases (see \citealt{gould2014a} for a complete discussion).
However, as shown in Figure \ref{fig:sigma_ratio_gamma_cdf}, the fractional precision of the vector proper motion from the high-resolution data is generally much better than that of $\gamma$.
I thus assume that the latter will set the minimum fractional precision of $M_{\ell}$ and show its distribution in Figure \ref{fig:sigma_ratio_gamma_cdf}.
Even taking this to be a conservative lower limit, only $\sim$14$\%$ of events would have $\sigma_{\gamma}/\gamma \leq 0.1$, and $\sim$40$\%$ would have $\sigma_{\gamma}/\gamma > 0.5$.
Nevertheless, photometric follow-up of microlensing events by groups such as the Microlensing Follow-Up Network \citep{gould2006} and RoboNet \citep{tsapras2009} provide high-cadence coverage that can increase $f$ in Equation (\ref{eq:sigma_gamma}) by up to an order of magnitude.

\acknowledgments
I graciously acknowledge B. Scott Gaudi for his support, patience, and insightful comments.
I also thank Matthew T.~Penny, David M.~Nataf, and Andrew P.~Gould for stimulating discussion, Susan Appel for her loving support, Richard Henderson for his inspiration, and Jennifer van Saders and Benjamin Shappee for use of their air mattress.
This material is based in part upon work supported by the National Science Foundation (NSF) Graduate Research Fellowship Program under Grant No.\ DGE-0822215, and an international travel allowance through the Graduate Research Opportunities Worldwide, taken to Cheongju, Korea.
Any opinions, findings, and conclusions or recommendations expressed in this material are those of the author and do not necessarily reflect the views of the NSF.
I recognize the direct support of The Ohio State University through a Distinguished University Fellowship.
This research has made use of the Exoplanet Orbit Database and the Exoplanet Data Explorer at exoplanets.org.


\begin{thebibliography}{}
\expandafter\ifx\csname natexlab\endcsname\relax\def\natexlab#1{#1}\fi

\bibitem[{{Atwood} {et~al.}(2012){Atwood}, {O'Brien}, {Colarosa}, {Mason},
  {Johnson}, {Pappalardo}, {Derwent}, {Schaller}, {Lee}, {Kim}, {Park}, {Cha},
  {Jorden}, {Darby}, {Walker}, \& {Renshaw}}]{atwood2012}
{Atwood}, B., {O'Brien}, T.~P., {Colarosa}, C., {et~al.} 2012, in Society of
  Photo-Optical Instrumentation Engineers (SPIE) Conference Series, Vol. 8446,
  Society of Photo-Optical Instrumentation Engineers (SPIE) Conference Series

\bibitem[{{Baraffe} {et~al.}(1998){Baraffe}, {Chabrier}, {Allard}, \&
  {Hauschildt}}]{baraffe1998}
{Baraffe}, I., {Chabrier}, G., {Allard}, F., \& {Hauschildt}, P.~H. 1998, \aap,
  337, 403

\bibitem[{{Baraffe} {et~al.}(2002){Baraffe}, {Chabrier}, {Allard}, \&
  {Hauschildt}}]{baraffe2002}
---. 2002, \aap, 382, 563

\bibitem[{{Batista} {et~al.}(2011){Batista}, {Gould}, {Dieters}, {Dong},
  {Bond}, {Beaulieu}, {Maoz}, {Monard}, {Christie}, {McCormick}, {Albrow},
  {Horne}, {Tsapras}, {Burgdorf}, {Calchi Novati}, {Skottfelt}, {Caldwell},
  {Koz{\l}owski}, {Kubas}, {Gaudi}, {Han}, {Bennett}, {An}, {MOA
  Collaboration}, {Abe}, {Botzler}, {Douchin}, {Freeman}, {Fukui}, {Furusawa},
  {Hearnshaw}, {Hosaka}, {Itow}, {Kamiya}, {Kilmartin}, {Korpela}, {Lin},
  {Ling}, {Makita}, {Masuda}, {Matsubara}, {Miyake}, {Muraki}, {Nagaya},
  {Nishimoto}, {Ohnishi}, {Okumura}, {Perrott}, {Rattenbury}, {Saito},
  {Sullivan}, {Sumi}, {Sweatman}, {Tristram}, {von Seggern}, {Yock}, {PLANET
  Collaboration}, {Brillant}, {Calitz}, {Cassan}, {Cole}, {Cook}, {Coutures},
  {Dominis Prester}, {Donatowicz}, {Greenhill}, {Hoffman}, {Jablonski}, {Kane},
  {Kains}, {Marquette}, {Martin}, {Martioli}, {Meintjes}, {Menzies},
  {Pedretti}, {Pollard}, {Sahu}, {Vinter}, {Wambsganss}, {Watson}, {Williams},
  {Zub}, {FUN Collaboration}, {Allen}, {Bolt}, {Bos}, {DePoy}, {Drummond},
  {Eastman}, {Gal-Yam}, {Gorbikov}, {Higgins}, {Janczak}, {Kaspi}, {Lee},
  {Mallia}, {Maury}, {Monard}, {Moorhouse}, {Morgan}, {Natusch}, {Ofek},
  {Park}, {Pogge}, {Polishook}, {Santallo}, {Shporer}, {Spector}, {Thornley},
  {Yee}, {MiNDSTEp Consortium}, {Bozza}, {Browne}, {Dominik}, {Dreizler},
  {Finet}, {Glitrup}, {Grundahl}, {Harps{\o}e}, {Hessman}, {Hinse},
  {Hundertmark}, {J{\o}rgensen}, {Liebig}, {Maier}, {Mancini}, {Mathiasen},
  {Rahvar}, {Ricci}, {Scarpetta}, {Southworth}, {Surdej}, {Zimmer}, {RoboNet
  Collaboration}, {Allan}, {Bramich}, {Snodgrass}, {Steele}, \&
  {Street}}]{batista2011}
{Batista}, V., {Gould}, A., {Dieters}, S., {et~al.} 2011, \aap, 529, A102

\bibitem[{{Batista} {et~al.}(2014){Batista}, {Beaulieu}, {Gould}, {Bennett},
  {Yee}, {Fukui}, {Gaudi}, {Sumi}, \& {Udalski}}]{batista2014}
{Batista}, V., {Beaulieu}, J.-P., {Gould}, A., {et~al.} 2014, \apj, 780, 54

\bibitem[{{Bennett} {et~al.}(2006){Bennett}, {Anderson}, {Bond}, {Udalski}, \&
  {Gould}}]{bennett2006}
{Bennett}, D.~P., {Anderson}, J., {Bond}, I.~A., {Udalski}, A., \& {Gould}, A.
  2006, \apjl, 647, L171

\bibitem[{{Bennett} {et~al.}(2007){Bennett}, {Anderson}, \&
  {Gaudi}}]{bennett2007}
{Bennett}, D.~P., {Anderson}, J., \& {Gaudi}, B.~S. 2007, \apj, 660, 781

\bibitem[{{Bond} {et~al.}(2001){Bond}, {Abe}, {Dodd}, {Hearnshaw}, {Honda},
  {Jugaku}, {Kilmartin}, {Marles}, {Masuda}, {Matsubara}, {Muraki}, {Nakamura},
  {Nankivell}, {Noda}, {Noguchi}, {Ohnishi}, {Rattenbury}, {Reid}, {Saito},
  {Sato}, {Sekiguchi}, {Skuljan}, {Sullivan}, {Sumi}, {Takeuti}, {Watase},
  {Wilkinson}, {Yamada}, {Yanagisawa}, \& {Yock}}]{bond2001}
{Bond}, I.~A., {Abe}, F., {Dodd}, R.~J., {et~al.} 2001, \mnras, 327, 868

\bibitem[{{Bond} {et~al.}(2004){Bond}, {Udalski}, {Jaroszy{\'n}ski},
  {Rattenbury}, {Paczy{\'n}ski}, {Soszy{\'n}ski}, {Wyrzykowski},
  {Szyma{\'n}ski}, {Kubiak}, {Szewczyk}, {{\.Z}ebru{\'n}}, {Pietrzy{\'n}ski},
  {Abe}, {Bennett}, {Eguchi}, {Furuta}, {Hearnshaw}, {Kamiya}, {Kilmartin},
  {Kurata}, {Masuda}, {Matsubara}, {Muraki}, {Noda}, {Okajima}, {Sako},
  {Sekiguchi}, {Sullivan}, {Sumi}, {Tristram}, {Yanagisawa}, {Yock}, \& {OGLE
  Collaboration}}]{bond2004}
{Bond}, I.~A., {Udalski}, A., {Jaroszy{\'n}ski}, M., {et~al.} 2004, \apjl, 606,
  L155

\bibitem[{{Bozza}(2000)}]{bozza2000}
{Bozza}, V. 2000, \aap, 355, 423

\bibitem[{{Cardelli} {et~al.}(1989){Cardelli}, {Clayton}, \&
  {Mathis}}]{cardelli1989}
{Cardelli}, J.~A., {Clayton}, G.~C., \& {Mathis}, J.~S. 1989, \apj, 345, 245

\bibitem[{{Cassan} {et~al.}(2012){Cassan}, {Kubas}, {Beaulieu}, {Dominik},
  {Horne}, {Greenhill}, {Wambsganss}, {Menzies}, {Williams}, {J{\o}rgensen},
  {Udalski}, {Bennett}, {Albrow}, {Batista}, {Brillant}, {Caldwell}, {Cole},
  {Coutures}, {Cook}, {Dieters}, {Prester}, {Donatowicz}, {Fouqu{\'e}}, {Hill},
  {Kains}, {Kane}, {Marquette}, {Martin}, {Pollard}, {Sahu}, {Vinter},
  {Warren}, {Watson}, {Zub}, {Sumi}, {Szyma{\'n}ski}, {Kubiak}, {Poleski},
  {Soszynski}, {Ulaczyk}, {Pietrzy{\'n}ski}, \& {Wyrzykowski}}]{cassan2012}
{Cassan}, A., {Kubas}, D., {Beaulieu}, J.-P., {et~al.} 2012, \nat, 481, 167

\bibitem[{{Chung} {et~al.}(2005){Chung}, {Han}, {Park}, {Kim}, {Kang}, {Ryu},
  {Kim}, {Jeon}, {Lee}, {Chang}, {Lee}, \& {Kang}}]{chung2005}
{Chung}, S.-J., {Han}, C., {Park}, B.-G., {et~al.} 2005, \apj, 630, 535

\bibitem[{{Dong} {et~al.}(2009){Dong}, {Gould}, {Udalski}, {Anderson},
  {Christie}, {Gaudi}, {OGLE Collaboration}, {Jaroszy{\'n}ski}, {Kubiak},
  {Szyma{\'n}ski}, {Pietrzy{\'n}ski}, {Soszy{\'n}ski}, {Szewczyk}, {Ulaczyk},
  {Wyrzykowski}, {{$\mu$}FUN Collaboration}, {DePoy}, {Fox}, {Gal-Yam}, {Han},
  {L{\'e}pine}, {McCormick}, {Ofek}, {Park}, {Pogge}, {MOA Collaboration},
  {Abe}, {Bennett}, {Bond}, {Britton}, {Gilmore}, {Hearnshaw}, {Itow},
  {Kamiya}, {Kilmartin}, {Korpela}, {Masuda}, {Matsubara}, {Motomura},
  {Muraki}, {Nakamura}, {Ohnishi}, {Okada}, {Rattenbury}, {Saito}, {Sako},
  {Sasaki}, {Sullivan}, {Sumi}, {Tristram}, {Yanagisawa}, {Yock}, {Yoshoika},
  {PLANET/RoboNet Collaborations}, {Albrow}, {Beaulieu}, {Brillant}, {Calitz},
  {Cassan}, {Cook}, {Coutures}, {Dieters}, {Prester}, {Donatowicz},
  {Fouqu{\'e}}, {Greenhill}, {Hill}, {Hoffman}, {Horne}, {J{\o}rgensen},
  {Kane}, {Kubas}, {Marquette}, {Martin}, {Meintjes}, {Menzies}, {Pollard},
  {Sahu}, {Vinter}, {Wambsganss}, {Williams}, {Bode}, {Bramich}, {Burgdorf},
  {Snodgrass}, {Steele}, {Doublier}, \& {Foellmi}}]{dong2009}
{Dong}, S., {Gould}, A., {Udalski}, A., {et~al.} 2009, \apj, 695, 970

\bibitem[{{Duquennoy} \& {Mayor}(1991)}]{duquennoy1991}
{Duquennoy}, A., \& {Mayor}, M. 1991, \aap, 248, 485

\bibitem[{{Erdl} \& {Schneider}(1993)}]{erdl1993}
{Erdl}, H., \& {Schneider}, P. 1993, \aap, 268, 453

\bibitem[{{Fischer} \& {Marcy}(1992)}]{fischer1992}
{Fischer}, D.~A., \& {Marcy}, G.~W. 1992, \apj, 396, 178

\bibitem[{{Gould}(2000)}]{gould2000}
{Gould}, A. 2000, \apj, 535, 928

\bibitem[{{Gould}(2014)}]{gould2014a}
---. 2014, ArXiv e-prints, arXiv:1408.0797

\bibitem[{{Gould} {et~al.}(2003){Gould}, {Gaudi}, \& {Han}}]{gould2003}
{Gould}, A., {Gaudi}, B.~S., \& {Han}, C. 2003, \apjl, 591, L53

\bibitem[{{Gould} {et~al.}(1994){Gould}, {Miralda-Escude}, \&
  {Bahcall}}]{gould1994}
{Gould}, A., {Miralda-Escude}, J., \& {Bahcall}, J.~N. 1994, \apjl, 423, L105

\bibitem[{{Gould} {et~al.}(2006){Gould}, {Udalski}, {An}, {Bennett}, {Zhou},
  {Dong}, {Rattenbury}, {Gaudi}, {Yock}, {Bond}, {Christie}, {Horne},
  {Anderson}, {Stanek}, {DePoy}, {Han}, {McCormick}, {Park}, {Pogge},
  {Poindexter}, {Soszy{\'n}ski}, {Szyma{\'n}ski}, {Kubiak}, {Pietrzy{\'n}ski},
  {Szewczyk}, {Wyrzykowski}, {Ulaczyk}, {Paczy{\'n}ski}, {Bramich},
  {Snodgrass}, {Steele}, {Burgdorf}, {Bode}, {Botzler}, {Mao}, \&
  {Swaving}}]{gould2006}
{Gould}, A., {Udalski}, A., {An}, D., {et~al.} 2006, \apjl, 644, L37

\bibitem[{{Gould} {et~al.}(2009){Gould}, {Udalski}, {Monard}, {Horne}, {Dong},
  {Miyake}, {Sahu}, {Bennett}, {Wyrzykowski}, {Soszy{\'n}ski}, {Szyma{\'n}ski},
  {Kubiak}, {Pietrzy{\'n}ski}, {Szewczyk}, {Ulaczyk}, {OGLE Collaboration},
  {Allen}, {Christie}, {DePoy}, {Gaudi}, {Han}, {Lee}, {McCormick}, {Natusch},
  {Park}, {Pogge}, {{$\mu$}FUN Collaboration}, {Allan}, {Bode}, {Bramich},
  {Burgdorf}, {Dominik}, {Fraser}, {Kerins}, {Mottram}, {Snodgrass}, {Steele},
  {Street}, {Tsapras}, {RoboNet Collaboration}, {Abe}, {Bond}, {Botzler},
  {Fukui}, {Furusawa}, {Hearnshaw}, {Itow}, {Kamiya}, {Kilmartin}, {Korpela},
  {Lin}, {Ling}, {Masuda}, {Matsubara}, {Muraki}, {Nagaya}, {Ohnishi},
  {Okumura}, {Perrott}, {Rattenbury}, {Saito}, {Sako}, {Skuljan}, {Sullivan},
  {Sumi}, {Sweatman}, {Tristram}, {Yock}, {MOA Collaboration}, {Albrow},
  {Beaulieu}, {Coutures}, {Calitz}, {Caldwell}, {Fouque}, {Martin}, {Williams},
  \& {PLANET Collaboration}}]{gould2009}
{Gould}, A., {Udalski}, A., {Monard}, B., {et~al.} 2009, \apjl, 698, L147

\bibitem[{{Gould} {et~al.}(2014){Gould}, {Udalski}, {Shin}, {Porritt},
  {Skowron}, {Han}, {Yee}, {Koz{\l}owski}, {Choi}, {Poleski}, {Wyrzykowski},
  {Ulaczyk}, {Pietrukowicz}, {Mr{\'o}z}, {Szyma{\'n}ski}, {Kubiak},
  {Soszy{\'n}ski}, {Pietrzy{\'n}ski}, {Gaudi}, {Christie}, {Drummond},
  {McCormick}, {Natusch}, {Ngan}, {Tan}, {Albrow}, {DePoy}, {Hwang}, {Jung},
  {Lee}, {Park}, {Pogge}, {Abe}, {Bennett}, {Bond}, {Botzler}, {Freeman},
  {Fukui}, {Fukunaga}, {Itow}, {Koshimoto}, {Larsen}, {Ling}, {Masuda},
  {Matsubara}, {Muraki}, {Namba}, {Ohnishi}, {Philpott}, {Rattenbury}, {Saito},
  {Sullivan}, {Sumi}, {Suzuki}, {Tristram}, {Tsurumi}, {Wada}, {Yamai}, {Yock},
  {Yonehara}, {Shvartzvald}, {Maoz}, {Kaspi}, \& {Friedmann}}]{gould2014b}
{Gould}, A., {Udalski}, A., {Shin}, I.-G., {et~al.} 2014, Science, 345, 46

\bibitem[{{Han}(2006)}]{han2006}
{Han}, C. 2006, \apj, 638, 1080

\bibitem[{{Han} \& {Gould}(1995{\natexlab{a}})}]{han1995a}
{Han}, C., \& {Gould}, A. 1995{\natexlab{a}}, \apj, 449, 521

\bibitem[{{Han} \& {Gould}(1995{\natexlab{b}})}]{han1995b}
---. 1995{\natexlab{b}}, \apj, 447, 53

\bibitem[{{Han} {et~al.}(2013){Han}, {Udalski}, {Choi}, {Yee}, {Gould},
  {Christie}, {Tan}, {Szyma{\'n}ski}, {Kubiak}, {Soszy{\'n}ski},
  {Pietrzy{\'n}ski}, {Poleski}, {Ulaczyk}, {Pietrukowicz}, {Koz{\l}owski},
  {Skowron}, {Wyrzykowski}, {OGLE Collaboration}, {Almeida}, {Batista},
  {Depoy}, {Dong}, {Drummond}, {Gaudi}, {Hwang}, {Jablonski}, {Jung}, {Lee},
  {Koo}, {McCormick}, {Monard}, {Natusch}, {Ngan}, {Park}, {Pogge}, {Porritt},
  {Shin}, \& {{$\mu$}FUN Collaboration}}]{han2013}
{Han}, C., {Udalski}, A., {Choi}, J.-Y., {et~al.} 2013, \apjl, 762, L28

\bibitem[{{Hardy} \& {Walker}(1995)}]{hardy1995}
{Hardy}, S.~J., \& {Walker}, M.~A. 1995, \mnras, 276, L79

\bibitem[{{Henderson} {et~al.}(2014{\natexlab{a}}){Henderson}, {Gaudi}, {Han},
  {Skowron}, {Penny}, {Nataf}, \& {Gould}}]{sexypants2014a}
{Henderson}, C.~B., {Gaudi}, B.~S., {Han}, C., {et~al.} 2014{\natexlab{a}},
  \apj, 794, 52

\bibitem[{{Henderson} {et~al.}(2014{\natexlab{b}}){Henderson}, {Park}, {Sumi},
  {Udalski}, {Gould}, {Tsapras}, {Han}, {Gaudi}, {Bozza}, {Abe}, {Bennett},
  {Bond}, {Botzler}, {Freeman}, {Fukui}, {Fukunaga}, {Itow}, {Koshimoto},
  {Ling}, {Masuda}, {Matsubara}, {Muraki}, {Namba}, {Ohnishi}, {Rattenbury},
  {Saito}, {Sullivan}, {Suzuki}, {Sweatman}, {Tristram}, {Tsurumi}, {Wada},
  {Yamai}, {Yock}, {Yonehara}, {MOA Collaboration}, {Szyma{\'n}ski}, {Kubiak},
  {Pietrzy{\'n}ski}, {Soszy{\'n}ski}, {Skowron}, {Koz{\l}owski}, {Poleski},
  {Ulaczyk}, {Wyrzykowski}, {Pietrukowicz}, {The OGLE Collaboration},
  {Almeida}, {Bos}, {Choi}, {Christie}, {Depoy}, {Dong}, {Friedmann}, {Hwang},
  {Jablonski}, {Jung}, {Kaspi}, {Lee}, {Maoz}, {McCormick}, {Moorhouse},
  {Natusch}, {Ngan}, {Pogge}, {Shin}, {Shvartzvald}, {Tan}, {Thornley}, {Yee},
  {The {$\mu$}FUN Collaboration}, {Allan}, {Bramich}, {Browne}, {Dominik},
  {Horne}, {Hundertmark}, {Figuera Jaimes}, {Kains}, {Snodgrass}, {Steele},
  {Street}, \& {The RoboNet Collaboration}}]{sexypants2014b}
{Henderson}, C.~B., {Park}, H., {Sumi}, T., {et~al.} 2014{\natexlab{b}}, \apj,
  794, 71

\bibitem[{{Holtzman} {et~al.}(1998){Holtzman}, {Watson}, {Baum}, {Grillmair},
  {Groth}, {Light}, {Lynds}, \& {O'Neil}}]{holtzman1998}
{Holtzman}, J.~A., {Watson}, A.~M., {Baum}, W.~A., {et~al.} 1998, \aj, 115,
  1946

\bibitem[{{Hwang} {et~al.}(2013){Hwang}, {Choi}, {Bond}, {Sumi}, {Han},
  {Gaudi}, {Gould}, {Bozza}, {Beaulieu}, {Tsapras}, {Abe}, {Bennett},
  {Botzler}, {Chote}, {Freeman}, {Fukui}, {Fukunaga}, {Harris}, {Itow},
  {Koshimoto}, {Ling}, {Masuda}, {Matsubara}, {Muraki}, {Namba}, {Ohnishi},
  {Rattenbury}, {Saito}, {Sullivan}, {Sweatman}, {Suzuki}, {Tristram}, {Wada},
  {Yamai}, {Yock}, {Yonehara}, {The MOA Collaboration}, {de Almeida}, {DePoy},
  {Dong}, {Jablonski}, {Jung}, {Kavka}, {Lee}, {Park}, {Pogge}, {Shin}, {Yee},
  {The {$\mu$}FUN Collaboration}, {Albrow}, {Bachelet}, {Batista}, {Brillant},
  {Caldwell}, {Cassan}, {Cole}, {Corrales}, {Coutures}, {Dieters}, {Dominis
  Prester}, {Donatowicz}, {Fouqu{\'e}}, {Greenhill}, {J{\o}rgensen}, {Kane},
  {Kubas}, {Marquette}, {Martin}, {Meintjes}, {Menzies}, {Pollard}, {Williams},
  {Wouters}, {The PLANET Collaboration}, {Bramich}, {Dominik}, {Horne},
  {Browne}, {Hundertmark}, {Ipatov}, {Kains}, {Snodgrass}, {Steele}, {Street},
  \& {The RoboNet Collaboration}}]{hwang2013}
{Hwang}, K.-H., {Choi}, J.-Y., {Bond}, I.~A., {et~al.} 2013, \apj, 778, 55

\bibitem[{{Ida} \& {Lin}(2005)}]{ida2005}
{Ida}, S., \& {Lin}, D.~N.~C. 2005, \apj, 626, 1045

\bibitem[{{Janczak} {et~al.}(2010){Janczak}, {Fukui}, {Dong}, {Monard},
  {Koz{\l}owski}, {Gould}, {Beaulieu}, {Kubas}, {Marquette}, {Sumi}, {Bond},
  {Bennett}, {Abe}, {Furusawa}, {Hearnshaw}, {Hosaka}, {Itow}, {Kamiya},
  {Korpela}, {Kilmartin}, {Lin}, {Ling}, {Makita}, {Masuda}, {Matsubara},
  {Miyake}, {Muraki}, {Nagaya}, {Nagayama}, {Nishimoto}, {Ohnishi}, {Perrott},
  {Rattenbury}, {Sako}, {Saito}, {Skuljan}, {Sullivan}, {Sweatman}, {Tristram},
  {Yock}, {MOA Collaboration}, {An}, {Christie}, {Chung}, {DePoy}, {Gaudi},
  {Han}, {Lee}, {Mallia}, {Natusch}, {Park}, {Pogge}, {{$\mu$}FUN
  Collaboration}, {Anguita}, {Calchi Novati}, {Dominik}, {J{\o}rgensen},
  {Masi}, {Mathiasen}, {MiNDSTEp Collaboration}, {Batista}, {Brillant},
  {Cassan}, {Cole}, {Corrales}, {Coutures}, {Dieters}, {Fouqu{\'e}},
  {Greenhill}, \& {PLANET Collaboration}}]{janczak2010}
{Janczak}, J., {Fukui}, A., {Dong}, S., {et~al.} 2010, \apj, 711, 731

\bibitem[{{Kappler} {et~al.}(2012){Kappler}, {Kappler}, {Poteet}, {Cauthen},
  {Park}, {Lee}, {Kim}, \& {Cha}}]{kappler2012}
{Kappler}, N., {Kappler}, L., {Poteet}, W.~M., {et~al.} 2012, in Society of
  Photo-Optical Instrumentation Engineers (SPIE) Conference Series, Vol. 8444,
  Society of Photo-Optical Instrumentation Engineers (SPIE) Conference Series

\bibitem[{{Kennedy} \& {Kenyon}(2008)}]{kennedy2008}
{Kennedy}, G.~M., \& {Kenyon}, S.~J. 2008, \apj, 673, 502

\bibitem[{{Kim} {et~al.}(2010){Kim}, {Park}, {Lee}, {Yuk}, {Han}, {O'Brien},
  {Gould}, {Lee}, \& {Kim}}]{kim2010}
{Kim}, S.-L., {Park}, B.-G., {Lee}, C.-U., {et~al.} 2010, in Society of
  Photo-Optical Instrumentation Engineers (SPIE) Conference Series, Vol. 7733,
  Society of Photo-Optical Instrumentation Engineers (SPIE) Conference Series

\bibitem[{{Kim} {et~al.}(2011){Kim}, {Park}, {Lee}, {Kappler}, {Kappler},
  {Poteet}, {Cauthen}, {Blanco}, {Buchroeder}, {Teran}, {Freestone}, {Lee},
  {Cho}, {Yuk}, {Chun}, {Jin}, \& {Cha}}]{kim2011}
{Kim}, S.-L., {Park}, B.-G., {Lee}, C.-U., {et~al.} 2011, in Society of
  Photo-Optical Instrumentation Engineers (SPIE) Conference Series, Vol. 8151,
  Society of Photo-Optical Instrumentation Engineers (SPIE) Conference Series

\bibitem[{{Krisciunas} \& {Schaefer}(1991)}]{krisciunas1991}
{Krisciunas}, K., \& {Schaefer}, B.~E. 1991, \pasp, 103, 1033

\bibitem[{{Lada}(2006)}]{lada2006}
{Lada}, C.~J. 2006, \apjl, 640, L63

\bibitem[{{Laney} {et~al.}(2012){Laney}, {Joner}, \&
  {Pietrzy{\'n}ski}}]{laney2012}
{Laney}, C.~D., {Joner}, M.~D., \& {Pietrzy{\'n}ski}, G. 2012, \mnras, 419,
  1637

\bibitem[{{Lissauer}(1987)}]{lissauer1987}
{Lissauer}, J.~J. 1987, \icarus, 69, 249

\bibitem[{{Majewski} {et~al.}(2011){Majewski}, {Zasowski}, \&
  {Nidever}}]{majewski2011}
{Majewski}, S.~R., {Zasowski}, G., \& {Nidever}, D.~L. 2011, \apj, 739, 25

\bibitem[{{McGregor} {et~al.}(2012){McGregor}, {Bloxham}, {Boz}, {Davies},
  {Doolan}, {Ellis}, {Hart}, {Jones}, {Luvaul}, {Nielsen}, {Parcell}, {Sharp},
  {Stevanovic}, \& {Young}}]{mcgregor2012}
{McGregor}, P.~J., {Bloxham}, G.~J., {Boz}, R., {et~al.} 2012, in Society of
  Photo-Optical Instrumentation Engineers (SPIE) Conference Series, Vol. 8446,
  Society of Photo-Optical Instrumentation Engineers (SPIE) Conference Series

\bibitem[{{Nataf} {et~al.}(2013){Nataf}, {Gould}, {Fouqu{\'e}}, {Gonzalez},
  {Johnson}, {Skowron}, {Udalski}, {Szyma{\'n}ski}, {Kubiak},
  {Pietrzy{\'n}ski}, {Soszy{\'n}ski}, {Ulaczyk}, {Wyrzykowski}, \&
  {Poleski}}]{nataf2013}
{Nataf}, D.~M., {Gould}, A., {Fouqu{\'e}}, P., {et~al.} 2013, \apj, 769, 88

\bibitem[{{Nidever} {et~al.}(2012){Nidever}, {Zasowski}, \&
  {Majewski}}]{nidever2012}
{Nidever}, D.~L., {Zasowski}, G., \& {Majewski}, S.~R. 2012, \apjs, 201, 35

\bibitem[{{Peale}(1997)}]{peale1997}
{Peale}, S.~J. 1997, \icarus, 127, 269

\bibitem[{{Poteet} {et~al.}(2012){Poteet}, {Cauthen}, {Kappler}, {Kappler},
  {Park}, {Lee}, {Kim}, \& {Cha}}]{poteet2012}
{Poteet}, W.~M., {Cauthen}, H.~K., {Kappler}, N., {et~al.} 2012, in Society of
  Photo-Optical Instrumentation Engineers (SPIE) Conference Series, Vol. 8444,
  Society of Photo-Optical Instrumentation Engineers (SPIE) Conference Series

\bibitem[{{Raghavan} {et~al.}(2010){Raghavan}, {McAlister}, {Henry}, {Latham},
  {Marcy}, {Mason}, {Gies}, {White}, \& {ten Brummelaar}}]{raghavan2010}
{Raghavan}, D., {McAlister}, H.~A., {Henry}, T.~J., {et~al.} 2010, \apjs, 190,
  1

\bibitem[{{Sumi} {et~al.}(2003){Sumi}, {Abe}, {Bond}, {Dodd}, {Hearnshaw},
  {Honda}, {Honma}, {Kan-ya}, {Kilmartin}, {Masuda}, {Matsubara}, {Muraki},
  {Nakamura}, {Nishi}, {Noda}, {Ohnishi}, {Petterson}, {Rattenbury}, {Reid},
  {Saito}, {Saito}, {Sato}, {Sekiguchi}, {Skuljan}, {Sullivan}, {Takeuti},
  {Tristram}, {Wilkinson}, {Yanagisawa}, \& {Yock}}]{sumi2003}
{Sumi}, T., {Abe}, F., {Bond}, I.~A., {et~al.} 2003, \apj, 591, 204

\bibitem[{{Sumi} {et~al.}(2010){Sumi}, {Bennett}, {Bond}, {Udalski}, {Batista},
  {Dominik}, {Fouqu{\'e}}, {Kubas}, {Gould}, {Macintosh}, {Cook}, {Dong},
  {Skuljan}, {Cassan}, {Abe}, {Botzler}, {Fukui}, {Furusawa}, {Hearnshaw},
  {Itow}, {Kamiya}, {Kilmartin}, {Korpela}, {Lin}, {Ling}, {Masuda},
  {Matsubara}, {Miyake}, {Muraki}, {Nagaya}, {Nagayama}, {Ohnishi}, {Okumura},
  {Perrott}, {Rattenbury}, {Saito}, {Sako}, {Sullivan}, {Sweatman}, {Tristram},
  {Yock}, {MOA Collaboration}, {Beaulieu}, {Cole}, {Coutures}, {Duran},
  {Greenhill}, {Jablonski}, {Marboeuf}, {Martioli}, {Pedretti}, {Pejcha},
  {Rojo}, {Albrow}, {Brillant}, {Bode}, {Bramich}, {Burgdorf}, {Caldwell},
  {Calitz}, {Corrales}, {Dieters}, {Dominis Prester}, {Donatowicz}, {Hill},
  {Hoffman}, {Horne}, {J{\o}rgensen}, {Kains}, {Kane}, {Marquette}, {Martin},
  {Meintjes}, {Menzies}, {Pollard}, {Sahu}, {Snodgrass}, {Steele}, {Street},
  {Tsapras}, {Wambsganss}, {Williams}, {Zub}, {PLANET Collaboration},
  {Szyma{\'n}ski}, {Kubiak}, {Pietrzy{\'n}ski}, {Soszy{\'n}ski}, {Szewczyk},
  {Wyrzykowski}, {Ulaczyk}, {OGLE Collaboration}, {Allen}, {Christie}, {DePoy},
  {Gaudi}, {Han}, {Janczak}, {Lee}, {McCormick}, {Mallia}, {Monard}, {Natusch},
  {Park}, {Pogge}, {Santallo}, \& {{$\mu$}FUN Collaboration}}]{sumi2010}
{Sumi}, T., {Bennett}, D.~P., {Bond}, I.~A., {et~al.} 2010, \apj, 710, 1641

\bibitem[{{Tsapras} {et~al.}(2009){Tsapras}, {Street}, {Horne}, {Snodgrass},
  {Dominik}, {Allan}, {Steele}, {Bramich}, {Saunders}, {Rattenbury}, {Mottram},
  {Fraser}, {Clay}, {Burgdorf}, {Bode}, {Lister}, {Hawkins}, {Beaulieu},
  {Fouqu{\'e}}, {Albrow}, {Menzies}, {Cassan}, \&
  {Dominis-Prester}}]{tsapras2009}
{Tsapras}, Y., {Street}, R., {Horne}, K., {et~al.} 2009, Astronomische
  Nachrichten, 330, 4

\bibitem[{{Udalski}(2003)}]{udalski2003}
{Udalski}, A. 2003, AcA, 53, 291

\bibitem[{{Yee} {et~al.}(2012){Yee}, {Shvartzvald}, {Gal-Yam}, {Bond},
  {Udalski}, {Koz{\l}owski}, {Han}, {Gould}, {Skowron}, {Suzuki}, {Abe},
  {Bennett}, {Botzler}, {Chote}, {Freeman}, {Fukui}, {Furusawa}, {Itow},
  {Kobara}, {Ling}, {Masuda}, {Matsubara}, {Miyake}, {Muraki}, {Ohmori},
  {Ohnishi}, {Rattenbury}, {Saito}, {Sullivan}, {Sumi}, {Suzuki}, {Sweatman},
  {Takino}, {Tristram}, {Wada}, {MOA Collaboration}, {Szyma{\'n}ski}, {Kubiak},
  {Pietrzy{\'n}ski}, {Soszy{\'n}ski}, {Poleski}, {Ulaczyk}, {Wyrzykowski},
  {Pietrukowicz}, {OGLE Collaboration}, {Allen}, {Almeida}, {Batista}, {Bos},
  {Christie}, {DePoy}, {Dong}, {Drummond}, {Finkelman}, {Gaudi}, {Gorbikov},
  {Henderson}, {Higgins}, {Jablonski}, {Kaspi}, {Manulis}, {Maoz}, {McCormick},
  {McGregor}, {Monard}, {Moorhouse}, {Mu{\~n}oz}, {Natusch}, {Ngan}, {Ofek},
  {Pogge}, {Santallo}, {Tan}, {Thornley}, {Shin}, {Choi}, {Park}, {Lee}, {Koo},
  \& {{$\mu$}FUN Collaboration}}]{yee2012}
{Yee}, J.~C., {Shvartzvald}, Y., {Gal-Yam}, A., {et~al.} 2012, \apj, 755, 102

\bibitem[{{Zheng} {et~al.}(2004){Zheng}, {Flynn}, {Gould}, {Bahcall}, \&
  {Salim}}]{zheng2004}
{Zheng}, Z., {Flynn}, C., {Gould}, A., {Bahcall}, J.~N., \& {Salim}, S. 2004,
  \apj, 601, 500

\end{thebibliography}
\bibliographystyle{apj}

\end{document}